\documentclass[aps,preprint,amsmath,amssymb,showpacs,amsfonts]{revtex4-1}
\usepackage{epsfig}
\usepackage{graphicx}
\usepackage{dcolumn}
\usepackage{bm}
\usepackage{amsthm}
\usepackage{setspace}

\allowdisplaybreaks[1]

\newcommand{\del}{\partial}

\newcommand{\OnH}{\xrightarrow{r=0}}
\newcommand{\calL}{\mathcal{L}}
\newcommand{\tr}{\operatorname{tr}}

\begin{document}

\title{Fluid/Gravity Correspondence, Local Wald Entropy Current\\ and Gravitational Anomaly}

\author{Shira Chapman}
\email{shirator@post.tau.ac.il}
\author{Yasha Neiman}
\email{yashula@gmail.com}
\author{Yaron Oz}
\email{yaronoz@post.tau.ac.il}
\affiliation{Raymond and Beverly Sackler School of Physics and Astronomy, Tel-Aviv University, Tel-Aviv 69978, Israel}

\date{\today}

\begin{abstract}
We propose, in the framework of the fluid/gravity correspondence, a definition for a local horizon entropy current for higher-curvature gravitational theories.
The current is well-defined to first order in fluid gradients for general gravity actions with an algebraic dependence on the Riemann tensor.
As a detailed example, we consider five-dimensional Einstein-Maxwell theory with a mixed gauge-gravitational Chern-Simons term.
In this theory, we construct the proposed entropy current on a charged black-brane background, and show that it has a non-negative divergence.
Moreover, a complete correspondence between the charged black-brane horizon's dynamics and the hydrodynamics of an anomalous four-dimensional field theory is established. Our proposed entropy current is then found to coincide with the entropy current of the anomalous field theory fluid.
\end{abstract}
\pacs{04.70.-s, 11.25.Tq, 47.10.ad}
\maketitle

\begin{spacing}{1.374}
    \tableofcontents
\end{spacing}


\section{Introduction}
The fluid/gravity correspondence relates field theory hydrodynamics to black hole (brane) dynamics.
In the hydrodynamic regime the system is in approximate thermal equilibrium in each local region, while the parameters of the equilibrium  vary slowly between one region and the next.
In a particular setup the fluid/gravity correspondence follows from  the AdS/CFT correspondence (for a review see \cite{Aharony:1999ti}), when considering the hydrodynamic regime of the field theory \cite{Bhattacharyya:2008jc}.
The essential ingredients needed to relate fluids to gravity are the existence of a horizon
in the gravitational background that is related to a thermal equilibrium state in the field theory, and a derivative expansion around it \cite{Damour,Eling:2009pb,Eling:2009sj}.
Thus, one can define the relation between fluids and gravity
on more general backgrounds, e.g. the Rindler geometry \cite{Bredberg:2011jq,Compere:2012mt,Eling:2012ni}.
One can also think about the fluid/gravity correspondence as an extension of black hole thermodynamics, where charges are upgraded into local currents, and the black hole entropy \cite{Bekenstein:1973ur,Hawking:1974sw} into a local entropy current.

The local entropy current in Einstein gravity $s^{\mu}$
is a vector density intrinsic to the horizon \cite{Bhattacharyya:2008xc}. It is directed along its generating light rays, and its flux through a spatial slice of the horizon gives the area of the slice.
In this work we will propose, in the framework of the fluid/gravity correspondence, a definition for a local horizon entropy current for higher-curvature gravitational theories. This should be a vector density $s^\mu$ on the horizon hypersurface, such that for stationary solutions its flux equals the total Wald entropy \cite{Wald:1993nt}, while for non-stationary solutions it satisfies an increase law $\del_\mu s^\mu \geq 0$.
Our definition of the entropy current is potentially ambiguous. However, for actions with an algebraic dependence on the curvature tensor, we will show that current is well-defined to first order in fluid gradients.

An interesting application is to a five-dimensional Einstein-Maxwell theory with
a gauge-gravitational Chern-Simons term. For completeness, we also include a pure gauge Chern-Simons term which was already studied in \cite{Eling:2010hu}.
In this setup, we will construct the proposed entropy current to first order in gradients on a charged black-brane background, and find that its divergence is positive.

This particular 5d theory has attracted much interest recently, because it implements chiral \cite{Erdmenger:2008rm,bat}
and mixed chiral-gravitational \cite{Amado:2011zx,Landsteiner:2011iq,Landsteiner:2011cp} anomalies in the dual four-dimensional field theory.
With these anomalies, it has been shown that new, non-dissipative, transport coefficients arise \cite{Son:2009tf,Neiman:2011}, which are responsible for the generation of current along magnetic fields and vortices in the fluid. A particular manifestation of these transport coefficients (the chiral magnetic effect \cite{Kharzeev:2007jp}) was suggested to exhibit charge separation \cite{Kharzeev:2010gr} and spin distribution \cite{KerenZur:2010zw} in heavy ion collisions \cite{Abelev:2009uh}.

To establish fully the fluid/gravity duality between the anomalous QFT hydrodynamics and Einstein-Maxwell theory with
a gauge-gravitational Chern-Simons term, we will analyze the horizon dynamics along the lines of \cite{Eling:2010hu}, and obtain the complete fluid equations. We then find that the entropy current as defined in the hydrodynamics coincides with our proposed Wald entropy current.

The paper is organized as follows.
In section \ref{sec:pre}, we review essential ingredients for the discussion, including null horizons, the Bekenstein-Hawking entropy current, hydrodynamics and Wald entropy.
In section \ref{sec:current}, we define our Wald entropy current and study its properties. In section \ref{sec:chern}, we consider the Wald entropy current of the 5d Einstein-Maxwell theory with a gravitational Chern-Simons term. In section \ref{sec:brane}, we establish the duality between the horizon dynamics in this theory and the hydrodynamics of an anomalous QFT.
In the appendices, we provide the detailed calculations behind the results of sections \ref{sec:chern}-\ref{sec:brane}.

\section{Preliminaries} \label{sec:pre}

\subsection{Null horizons} \label{sec:pre:notations}

We will be interested in event horizons, or more generally in null hypersurfaces. In the following, we briefly review some of the required ingredients for our studies.
The general arguments are independent of the dimension, but for concreteness we will refer to a four-dimensional horizon embedded in a five-dimensional bulk
spacetime. This is also the relevant case for the $\mathrm{AdS}_5/\mathrm{CFT}_4$ correspondence, and for the particular model in sections \ref{sec:chern}-\ref{sec:brane}. We denote bulk coordinates by $x^A$, with capital Latin indices. Where relevant, we work in a coordinate system adapted to the horizon, i.e. we decompose $x^A = (r,x^\mu)$, with the horizon situated at $r = 0$. Small Greek indices denote tensors intrinsic to the horizon hypersurface.

The bulk metric is $g_{AB}$. The tangent covector to the horizon is denoted by $\ell_A = (\ell_r, 0_\mu)$. Since $\ell_A$ is null, it has no canonical normalization. However, we can always match its normalization with the choice of coordinates, so that we have $\ell_r = 1$. Raising the index with the inverse metric, we get the normal vector $\ell^A = g^{AB}\ell_B$. Being null, it is also tangent to the horizon, pointing along its generating lightrays. Therefore, we can restrict the 5d index, and write $\ell^\mu$ as an intrinsic 4d vector. Once again, this vector does not have a natural normalization. It is easy to see that the above conventions fix the following components of the inverse metric:
\begin{align}
 g^{rr} = 0; \quad g^{r\mu} = \ell^\mu \ .
\end{align}

The bulk metric induces a rich and peculiar geometry on the null horizon. For now, we only mention the key concept of \emph{surface gravity}. Given a choice of scaling for $\ell^A$ at each horizon point $x^\mu$, it can be shown that the covariant derivative $\ell^B\nabla_B\ell^A$ is again directed along $\ell^A$. The proportionality coefficient is the surface gravity $\kappa$:
\begin{align}
 \ell^B\nabla_B\ell^A = \kappa\ell^A \ .
\end{align}
We stress that $\kappa$ changes under local rescalings of $\ell^A$. In black hole thermodynamics, $\kappa = \nolinebreak2\pi T$ defines the horizon's temperature, and is taken to be non-negative.

\subsection{Bekenstein-Hawking entropy current}\label{sec:pre:BH}

In this subsection, we briefly review the concept of a local entropy current in Einstein gravity. We start with the Bekenstein-Hawking formula for the entropy of a horizon in equilibrium \cite{Bekenstein:1973ur,Hawking:1974sw}:
\begin{align}
 S = \frac{1}{4}A \ . \label{eq:Bekenstein}
\end{align}
This global area-proportional entropy can always be upgraded into a local current. Indeed, any null hypersurface is home to a canonical \emph{area current} $J_{\mathrm{Area}}^\mu$. This is a vector density intrinsic to the horizon and directed along its generating lightrays. As the name suggests, its flux through a spatial slice of the horizon gives the area of the slice.

$J_{\mathrm{Area}}^\mu$ can be defined as follows \cite{Bhattacharyya:2008xc}. Start with the spacetime volume density $\sqrt{-g}$, and dualize it to obtain the volume form $\sqrt{-g}\epsilon_{ABCDE}$ (we consider 5d spacetime as a generic example). Now take the pullback of 4 indices into the horizon and contract them with $\epsilon^{\mu\nu\rho\sigma}/4!$, where $\epsilon^{\mu\nu\rho\sigma}$ is the horizon's Levi-Civita density. The result is a quantity $\Lambda_A$ with unit 4d density weight, directed along the tangent covector $\ell_A$. Let us now raise its index with the spacetime metric. The resulting quantity $\Lambda^A$ is directed along the horizon's normal. For a non-null hypersurface, at this point we could have written $\Lambda^A = \sqrt{\gamma}n^A$, with $n^A$ the unit normal, thus defining the hypersurface volume density $\sqrt{\gamma}$. On a null horizon, however, there is no preferred unit normal. On the other hand, the normal $\Lambda^A$ is now \emph{tangent} to the horizon, pointing along its generating lightrays. Thus, we can interpret $\Lambda^A$ directly as an intrinsic vector density $J_{\mathrm{Area}}^\mu$ on the horizon.

By an immediate generalization of \eqref{eq:Bekenstein}, we can now define a local entropy current as:
\begin{align}
 s^\mu = \frac{1}{4}J_{\mathrm{Area}}^\mu \ . \label{eq:Bekenstein_current}
\end{align}
Does this current satisfy a local increase law $\del_\mu s^\mu \geq 0$? In general, there is no easy answer. The quantity $\del_\mu J_{\mathrm{Area}}^\mu$ is governed by the twice-projected component $E_{AB}\ell^A\ell^B$ of the Einstein equations. This component is known as the Focusing Equation. When decomposed in terms of horizon quantities, it reads:
\begin{align}
  \kappa\theta = \ell^\mu\del_\mu\theta + \frac{1}{3}\theta^2 + \sigma_{(H)}^2 + 8\pi\mathcal{T}_{\mu\nu}\ell^\mu\ell^\nu \ . \label{eq:focusing}
\end{align}
Here, $\mathcal{T}_{AB}$ is the bulk energy-momentum tensor; $\sigma^2_{(H)}$ is the (non-negative) square of the horizon's shear tensor; $\theta = \del_\mu J_{\mathrm{Area}}^\mu/v$ is the area expansion rate. The ``area density'' $v$ is the coefficient in the collinearity relation $J_{\mathrm{Area}}^\mu = v\ell^\mu$; in horizon-adapted coordinates, it's given by $v = \sqrt{-g}$.

We would like the area production rate $\del_\mu s^\mu = v\theta/4$ to be non-negative. As with the global entropy increase $\Delta S \geq 0$  \cite{Hawking:1971tu}, we see from \eqref{eq:focusing} that the answer depends on the null energy condition $\mathcal{T}_{AB}\ell^A\ell^B \geq 0$. However, this is not enough, since the second-derivative term $\ell^\mu\del_\mu\theta$ can have either sign. This term is related to the fact that an event horizon is defined not locally, but by boundary conditions at future infinity which ensure that the horizon is eternal. Indeed, these boundary conditions also play a role in the global area-increase theorem.

We conclude that in Einstein gravity, a natural entropy current is always defined, but a local entropy increase law $\del_\mu s^\mu \geq 0$ is not always satisfied. In the next subsection, we will review an important limit where this law \emph{does} hold. This (hydrodynamic) limit will occupy us for the bulk of the paper.

\subsection{The hydrodynamic ansatz in gravity} \label{sec:pre:hydro}

In non-gravitational systems, an entropy current can be defined in the limit described by hydrodynamics. This means that the system is in approximate thermal equilibrium in each local region, while the equilibrium parameters slowly vary between one region and the next. These local parameters can be encoded by a 4-velocity $u^\mu$, an entropy density $s$, and optionally also conserved charge densities $n^a$. One can also define the energy density $\epsilon$, the pressure $p$, the temperature $T$ and the chemical potentials $\mu_a$. The various thermal parameters are related by a material-specific equation of state $\epsilon(s,n^a)$, as well as by universal thermodynamic identities.

A fluid state necessarily requires a separation of scales. On the microscopic side, we have the scale determined by the temperature and the underlying dynamics, while on the macroscopic side we have the gradients of the equilibrium parameters. In gravity, such a separation of scales corresponds to a very large and homogeneous horizon, as compared to its surface gravity and the full spacetime curvature. Such horizons are easily constructed as planar black holes in AdS. Other examples are also conceivable, e.g. a Rindler horizon in Minkowski space.

To begin with, let us write down a gravitational ansatz for a horizon in \emph{global} equilibrium, satisfying the required separation of scales. We can do so using a constant Lorentzian metric $h_{\mu\nu}$ and a 4-velocity vector $u^\mu$, with $u_\mu u^\mu = -1$. Here and below, we implicitly raise and lower 4d indices with $h_{\mu\nu}$, rather than with the gravitational metric $g_{AB}$. The ansatz for the gravitational metric in a neighborhood of the horizon reads:
\begin{align}
 g_{rr} = 0;\quad g_{r\mu} = -u_\mu;\quad g^{(0)}_{\mu\nu} = f(r)P_{\mu\nu} + k(r)u_\mu u_\nu \ , \label{eq:g_0}
\end{align}
where $P_{\mu\nu} = h_{\mu\nu} + u_\mu u_\nu$ is the projector orthogonal to $u^\mu$. The coefficient functions $f(r)$ and $k(r)$ are constrained by the field equations. Note that with general $f(r)$ and $k(r)$, the ansatz \eqref{eq:g_0} holds not just for Einstein gravity, but for any gravitational action which allows fluid-like solutions. To describe more general fluids, we may add bulk fields other than the metric. Indeed, we will consider a bulk Maxwell field in sections \ref{sec:chern}-\ref{sec:brane}. The general form \eqref{eq:g_0} of the metric ansatz is again indifferent to these details. The ``$(0)$'' superscript on $g_{\mu\nu}$ stands for ``zeroth order'', and anticipates corrections.

The horizon of the solution \eqref{eq:g_0} is located at the value of $r$ where the function $k$ vanishes. Without loss of generality, we may take this value to be $r = 0$. The homogeneity of the solution implies that $r = 0$ is in fact a Killing horizon. Its Killing normal is given by:
\begin{align}
 \ell^A = (0, u^\mu); \quad \ell_A = (1, 0_\mu) \ . \label{eq:ell}
\end{align}
This normalization of $\ell^A$ is in accordance with the conventions of section \ref{sec:pre:notations}.

The velocity $u^\mu$ represents the rest frame of the black hole. The nondegenerate metric $h_{\mu\nu}$ can be interpreted as the metric in which the dual fluid lives. In the AdS/CFT context, it corresponds to the metric on the boundary, where the CFT is defined. The horizon value $f(0)$ is related to the horizon's area density (recall that $k(0)$ vanishes):
\begin{align}
 J_{\mathrm{Area}}^\mu = \sqrt{-h}f(0)^{3/2}u^\mu \ .
\end{align}
Thus, the entropy current in Einstein gravity reads:
\begin{align}
 s^\mu = \frac{1}{4}J_{\mathrm{Area}}^\mu = \frac{1}{4}\sqrt{-h}f(0)^{3/2}u^\mu \equiv \sqrt{-h}su^\mu \ , \label{eq:f_s_current}
\end{align}
and we see that $f(0)$ is related to the entropy density $s$:
\begin{align}
 f(0) = (4s)^{2/3} \ . \label{eq:f_s}
\end{align}
Similarly, the radial derivative $k'(0)$ is proportional to the horizon's surface gravity, and thus to its temperature:
\begin{align}
 k'(0) = -2\kappa = -4\pi T \ . \label{eq:temperature_def}
\end{align}
Unlike the relations \eqref{eq:f_s_current}-\eqref{eq:f_s} between area and entropy, eq. \eqref{eq:temperature_def} holds also for higher-curvature gravitational actions. 

Let us now move on from global equilibrium to one that is local in $x^\mu$. We take $h_{\mu\nu}$, $u^\mu$, $f$ and $k$ in \eqref{eq:g_0} to depend on $x^\mu$ with very small $\del_\mu$ derivatives. Note that this is in addition to the explicit dependence of $f$ and $k$ on the radial coordinate $r$, with derivatives which cannot be treated as small. Every quantity can now be expanded in powers of the small derivatives $\del_\mu$. We will refer to this power as the ``order'' of the quantity.

After introducing this dependence on $x^\mu$, the ansatz \eqref{eq:g_0} in general no longer solves the field equations. However, it still solves them at leading order. At higher orders, we must introduce corrections. By an appropriate choice of coordinates, we can keep the horizon at $r = 0$, and also prevent corrections to $g_{rr}$ and $g_{r\mu}$.
For $g_{\mu\nu}$, we will consider only first-order corrections $g^{(1)}_{\mu\nu}$. Since we never use the explicit form of $g^{(1)}_{\mu\nu}$ in this paper, there is no loss of generality involved: $g^{(1)}_{\mu\nu}$ can be understood to contain higher-order corrections as well.
Finally, we fix the velocity variable $u^\mu$ such that \eqref{eq:ell} holds exactly, i.e. $u^\mu$ always points along the horizon's null generators. This constrains the metric correction on the horizon:
\begin{align}
 r = 0:\quad g^{(1)}_{\mu\nu}u^\nu = 0 \ .
\end{align}

In the hydrodynamic limit defined above, the entropy current \eqref{eq:Bekenstein_current} of Einstein gravity \emph{does} have a local non-negative divergence.
To see this, let us consider the focusing equation \eqref{eq:focusing} order by order. Note that the horizon's shear and expansion parameters $\sigma_{(H)}$, $\theta$ necessarily contain at least one $\del_\mu$ gradient. Then at first order, eq. \eqref{eq:focusing} reads:
\begin{align} \label{eq:focusing_1}
  \kappa^{(0)}\theta^{(1)} = 8\pi\mathcal{T}^{(1)}_{AB}\ell^A\ell^B = 0 \ .
\end{align}
The last equality follows from our null-energy assumption $\mathcal{T}_{AB}\ell^A\ell^B \geq 0$, since it's impossible to write a non-vanishing positive-definite expression with one $\del_\mu$ gradient. Thus, at first order the entropy production rate $\del_\mu s^\mu \sim \theta$ vanishes. This is always the case with leading-order hydrodynamics, known as the ``ideal fluid'' limit. At second order, the focusing equation reads, using $\theta^{(1)} = 0$:
\begin{align} \label{eq:focusing_2}
 \kappa^{(0)}\theta^{(2)} = (\sigma_{(H)}^{(1)})^2 + 8\pi\mathcal{T}^{(2)}_{AB}\ell^A\ell^B \geq 0 \ .
\end{align}
As promised, at the leading nontrivial order, we see a non-negative local entropy production rate $\del_\mu s^\mu \sim \theta$. The problematic term $\ell^\mu\del_\mu\theta$ fell away due to the small-derivative expansion. In general, it's not guaranteed that the next-order contributions to $\del_\mu s^\mu$ will also be non-negative. However, in AdS/CFT, where the metric corrections are controlled by the AdS boundary conditions, positivity does hold at the next order; see e.g. \cite{Loganayagam:2008is}.

\subsection{Wald entropy}

Wald proposed a definition \cite{Wald:1993nt} for a notion of horizon entropy in general theories of gravity. The definition applies to stationary black holes in equilibrium, where the horizon's normal $\ell^A$ (with a suitable choice of normalization) is part of a Killing vector field. The entropy is identified with the Noether charge associated with transport along this Killing field. The local nature of diffeomorphism invariance then allows us to express this quantity as a local integral over a spatial slice of the horizon.

In the present paper, we consider gravity coupled to gauge fields. We limit ourselves to Lagrangians of the form $\calL(g_{AB}, R_{ABCD}, A_A, F_{AB})$, with an arbitrary \emph{algebraic} dependence on the Riemann tensor and the field strength. We do not exclude direct dependence on the gauge potential $A_A$, to allow for Chern-Simons terms. In this setup, the rank-2 ``Noether potential'' $Q^{AB}$ for diffeomorphisms along $\ell^A$ is given by \cite{Jacobson:1993vj,LopesCardoso:1999cv}:
\begin{align}
 Q^{AB} = -2\calL^{ABCD}\nabla_C\ell_D + 4\ell_D\nabla_C\calL^{ABCD} \ , \label{eq:Q}
\end{align}
where $\calL^{ABCD} \equiv \del\calL/\del R_{ABCD}$ is the derivative of the Lagrangian with respect to the Riemann tensor. $Q^{AB}$ is antisymmetric and carries unit density weight. To be precise, in the presence of gauge fields, \eqref{eq:Q} is the Noether potential for ``gauge-covariant'' diffeomorphisms, where together with $x^A \rightarrow x^A + \alpha\ell^A$ we perform the gauge transformation $A_A \rightarrow A_A - \del_A(\alpha\ell^B A_B)$. If we considered instead ``pure'' diffeomorphisms, \eqref{eq:Q} would have contained an additional gauge-dependent term of the form $\ell^C A_C\cdot\del\calL/\del F_{AB}$.

The horizon entropy is now given by an integral of $Q^{AB}$ over a spatial slice of the horizon:
\begin{align}
 S = \frac{2\pi}{\kappa}\int{Q^{AB}d\Sigma_{AB}} = \frac{1}{T}\int{Q^{AB}d\Sigma_{AB}} \ , \label{eq:S}
\end{align}
where $\kappa$ is the surface gravity associated with the horizon normal $\ell^A$. In Einstein gravity, we have $\calL^{ABCD} = (\sqrt{-g}/32\pi)(g^{AC}g^{BD} - g^{AD}g^{BC})$, and eq. \eqref{eq:S} recovers the Bekenstein-Hawking entropy \eqref{eq:Bekenstein}. We note that when calculating the entropy of stationary horizons, the second term in \eqref{eq:Q} doesn't contribute to the integral \cite{Jacobson:1993vj}, and may be omitted.
However, as we'll see in section \ref{sec:chern}, this term \emph{is} important for the positive divergence of a local entropy current in non-stationary solutions.

\section{Wald entropy current} \label{sec:current}

\subsection{Definition and possible ambiguities} \label{sec:current:definition}

We would like to define a local current for the Wald entropy \eqref{eq:S}. More precisely, we're looking for a vector density $s^\mu$ on the horizon hypersurface, such that for stationary solutions its flux equals the total entropy, while for non-stationary solutions it satisfies $\del_\mu s^\mu \geq 0$. The expression \eqref{eq:S} for the total entropy suggests an obvious candidate, which is the definition we adopt:
\begin{align}
 s^\mu = \frac{2\pi}{\kappa}Q^{\mu r} \ . \label{eq:s_short}
\end{align}
This formula is written in the horizon-adapted coordinates of section \ref{sec:pre:notations}. In coordinate-free terms, $Q^{\mu r}$ is the flux of the 5d antisymmetric density $Q^{AB}$ through the horizon. The result is a vector density on the horizon, with unit 4d density weight. In form language, this operation of taking the flux can be described in three steps:
\begin{enumerate}
 \item Dualize $Q^{AB}$ into a 3-form.
 \item Pullback this form into the horizon.
 \item Dualize back within the horizon geometry, resulting in a rank-1 (vector) density.
\end{enumerate}
Thus, we've defined an intermediate step between the Noether potential $Q^{AB}$ and the global entropy \eqref{eq:S}. In \eqref{eq:S}, a scalar density was obtained by taking the flux of $Q^{AB}$ through a codimension-2 surface (the spatial slice $\Sigma$). In our definition \eqref{eq:s_short}, we obtain a vector density by taking the flux through the codimension-1 horizon. It's clear that for a stationary solution,
taking the flux again with respect to the remaining index will bring us back to the scalar entropy \eqref{eq:S}.

The definition \eqref{eq:s_short} can be considered for any gravitational action. Here, we will work out its consequences for ``algebraic'' Lagrangians of the form $\calL(g_{AB}, R_{ABCD}, A_A, F_{AB})$, where $Q^{AB}$ is given by the formula \eqref{eq:Q}.
Using \eqref{eq:Q} and decomposing into radial and tangential components, \eqref{eq:s_short} can be written more explicitly as:
\begin{align}
  s^\mu = \frac{2\pi}{\kappa}\left(-2\calL^{\mu r\nu r}(\nabla_\nu\ell_r - \nabla_r\ell_\nu) + 4\ell_r\nabla_\nu\calL^{\mu r\nu r}\right) \ . \label{eq:s}
\end{align}
To arrive at this simplified expression, we used the antisymmetry $\calL^{ABCD} = -\calL^{ABDC}$ and the relations $\ell_\mu = 0$ and $\nabla_{[\mu}\ell_{\nu]} = 0$ on the horizon. The former relation is due to the horizon-adapted coordinates, while the latter follows from Frobenius' theorem $\ell_{[A}\nabla_B\ell_{C]} = 0$. We left the magnitude of $\ell_r$ in \eqref{eq:s} arbitrary, to exhibit explicitly the dependence on the scaling of $\ell_A$ without entangling it with the choice of coordinates.

For non-equilibrium solutions, the definition \eqref{eq:s_short}-\eqref{eq:s} is ambiguous, and we do not see a way to resolve the ambiguity generically. For Einstein gravity however, this can be done, as we will describe in section \ref{sec:current:Einstein}.
In higher-curvature theories, we must restrict ourselves to the hydrodynamic limit. There, at least for actions with an algebraic dependence on the $R_{ABCD}$, the ambiguities can be resolved at the first non-trivial order in small gradients.

The reason for the ambiguity is that we no longer have a Killing vector field. Therefore, we must ask ourselves what exactly is meant by $\ell^A$ and $\kappa$ in \eqref{eq:s}. It is natural to insist that $\ell^A$ remains normal to the horizon, and that $\kappa$ remains the surface gravity associated with $\ell^A$. It remains to determine the local scaling of $\ell^A$ on the horizon $r = 0$, as well as its behavior at $r\neq 0$. The latter is relevant due to the $\nabla_r\ell_\nu$ contribution to the first term in \eqref{eq:s}. For an unambiguous definition of $s^\mu$, we must either find a way to fix these freedoms or show that they are irrelevant.

We note that the definition \eqref{eq:s} is invariant under global rescalings of $\ell^A$: the explicit factor of $\ell^A$ is canceled against the implicit factor in $\kappa$ in the denominator. This was of course already necessary for an unambiguous definition of the global entropy. Thus, any possible ambiguities arise from derivatives of $\ell^A$. This includes the explicit derivatives in the parentheses, as well as the implicit derivative contained in $\kappa$.

\subsection{Weak Killing condition} \label{sec:current:Einstein}

As a necessary prerequisite, let us now ensure that our definition \eqref{eq:s} coincides with the Bekenstein current \eqref{eq:Bekenstein_current} for Einstein gravity. In the process, we will fix some of the freedom in the choice of $\ell^A$. Having convinced ourselves in the previous subsection that $s^\mu$ is invariant under global rescalings of $\ell^A$, we use the coordinate freedom to set $\ell_r = 1$,
 and look for ambiguities only in derivatives of $\ell^A$.

For the Einstein-Hilbert Lagrangian, we have:
\begin{align}
 \calL^{\mu r\nu r} &= \frac{\sqrt{-g}}{32\pi}(g^{\mu\nu}g^{rr} - g^{\mu r}g^{r\nu}) = \frac{\sqrt{-g}}{32\pi}(0 - \ell^\mu\ell^\nu) = -\frac{\sqrt{-g}}{32\pi}\ell^\mu\ell^\nu \ , \\
 \nabla_\nu\calL^{\mu r\nu r} &= 0 \ , \\
 s^\mu &= \frac{1}{8\kappa}\sqrt{-g}\ell^\mu\ell^\nu(\nabla_\nu\ell_r - \nabla_r\ell_\nu) \ . \label{eq:s_Einstein}
\end{align}
Here we encounter a problem. For stationary solutions, the $\nabla_r\ell_\nu$ derivative in \eqref{eq:s_Einstein} could be related to the $\nabla_\nu\ell_r$ derivative via the Killing condition $\nabla_A\ell_B = -\nabla_B\ell_A$. To make the expression \eqref{eq:s_Einstein} unambiguous, we must find a substitute for this constraint.

At each point on the horizon, we can fix one degree of freedom for the local scaling of $\ell^A$, plus 5 degrees of freedom for the radial derivative $\nabla_r\ell_A$. Of course, these are not enough to force the full Killing condition $\nabla_A\ell_B = -\nabla_B\ell_A$. They \emph{are} enough to force $\nabla_r\ell_\nu = -\nabla_\nu\ell_r$, but unfortunately this condition is not covariant: the restriction of a 5d lower index to the value $r$ depends on the choice of coordinates outside the horizon. The way out is to force the following covariant condition, given by contracting the Killing condition with $\ell^B$:
\begin{align}
 \ell^B(\nabla_A\ell_B + \nabla_B\ell_A) = 0 \ . \label{eq:killing_weak}
\end{align}
The $\mu$-component of this condition is in fact an identity:
\begin{align}
 \ell^B(\nabla_\mu\ell_B + \nabla_B\ell_\mu) = \frac{1}{2}\nabla_\mu(\ell_B\ell^B) + \kappa\ell_\mu = 0 + 0 = 0 \ ,
\end{align}
while the $r$-component takes the form:
\begin{align}
 \ell^\mu(\nabla_r\ell_\mu + \nabla_\mu\ell_r) = 0 \ .
\end{align}
Thus, our condition eats up just one degree of freedom, relating $\nabla_r(\ell_A\ell^A)$ with the surface gravity. Eq. \eqref{eq:s_Einstein} now becomes:
\begin{align}
 s^\mu = \frac{1}{4\kappa}\sqrt{-g}\ell^\mu\ell^\nu\nabla_\nu\ell_r = \frac{1}{4\kappa}\sqrt{-g}\ell^\mu\kappa\ell_r = \frac{1}{4}\sqrt{-g}\ell^\mu \ .
\end{align}
In our horizon-adapted coordinates, this coincides with the standard definition \eqref{eq:Bekenstein_current}.

\subsection{Non-ambiguity at first order} \label{sec:current:condition}

Returning to higher-curvature theories, let us consider the entropy current \eqref{eq:s} in the hydrodynamic limit described in section \ref{sec:pre:hydro}. The current and its ambiguities can now be examined order by order in gradients. We can expect better behavior in this limit than in the general case, because $\ell^A$ is now \emph{almost} a Killing vector.

At zeroth order, $\ell^A$ \emph{is} a Killing vector, and therefore $s^{\mu(0)}$ is well-defined. Now, the only available 4d vector at zeroth order is $u^\mu$. Thus, $s^{\mu(0)}$ is always directed along $u^\mu$, i.e. along the horizon's null generators. The only nontrivial content of $s^{\mu(0)}$ is therefore its magnitude, which is already captured by the scalar area density $Q^{AB}d\Sigma_{AB}$. This result is in line with the general rule that the leading-order hydrodynamics is completely determined by the \emph{thermo}dynamics. Another point of view is that at zeroth order, our Wald entropy current is simply a rescaled version of the Bekenstein current \eqref{eq:Bekenstein_current}, which always points along the null generators.

Thus, the first-order entropy current $s^{\mu(1)}$ is where we should start looking for nontrivial local information. Here the issue of ambiguities becomes relevant. Once again, we set $\ell_r = 1$. Expanding each piece of \eqref{eq:s} in small gradients, the ambiguous contribution at first order reads:
\begin{align}
 s_{\mathrm{ambig.}}^{\mu(1)} ={}& 2\pi\left(-2\calL^{\mu r\nu r(0)}\left(\frac{1}{\kappa}(\nabla_\nu\ell_r - \nabla_r\ell_\nu)\right)^{(1)}
  + 4\left(\frac{1}{\kappa}\right)^{(1)}\nabla_\nu\calL^{\mu r\nu r(0)} \right) \ . \label{eq:s_ambig}
\end{align}
It is known \cite{Jacobson:1993vj} that for stationary horizons, the $\nabla_\nu\calL^{\mu r\nu r}$ term (more generally, the term with no factors of $\nabla_{[A}\ell_{B]}$) vanishes. The proof is based on the existence of a bifurcation surface in the horizon's past \cite{Racz:1992bp}. Now, the zeroth order in fluid gradients is locally the same as the stationary solution, so for non-stationary solutions we still have $\nabla_\nu\calL^{\mu r\nu r(0)} = 0$. Therefore, the second term in \eqref{eq:s_ambig} doesn't contribute. It remains to consider the first term. This requires us to discuss the zeroth-order quantity $\calL^{\mu r\nu r(0)}$. The only available rank-2 tensors at zeroth order are $u^\mu u^\nu$ and the transverse projector $P^{\mu\nu}$. Thus, $\calL^{\mu r\nu r(0)}$ must take the form:
\begin{align}
 \calL^{\mu r\nu r(0)} = C_1 u^\mu u^\nu + C_2 P^{\mu\nu} \ , \label{eq:L_0}
\end{align}
The longitudinal coefficient $C_1$ determines the Wald entropy of a stationary solution. The coefficient $C_2$ can be shown to vanish, using $\nabla_\nu\calL^{\mu r\nu r(0)} = 0$. To demonstrate this, we will need the zeroth-order Christoffel symbols at $r = 0$ derived from the metric \eqref{eq:g_0}. The non-vanishing Christoffel symbols read:
\begin{align}\label{eq:Christoffel_0}
 \Gamma^{r(0)}_{r\mu} = -\frac{1}{2}k' u_\mu;\quad \Gamma^{\mu(0)}_{r\nu} = \frac{f'}{2f}P^\mu_\nu;\quad
 \Gamma^{\mu(0)}_{\nu\rho} = -\frac{1}{2}u^\mu(f' P_{\nu\rho} + k' u_\nu u_\rho) \ .
\end{align}
Note that the Christoffel symbols arise from the \emph{radial} derivatives of the coefficient functions $f$ and $k$, which are not small. The vanishing condition for $\nabla_\nu\calL^{\mu r\nu r(0)}$ can now be written as:
\begin{align}
 \begin{split}
   0 &= \nabla_\nu\calL^{\mu r\nu r(0)} \\
     &= \Gamma^{\mu(0)}_{\nu A}\calL^{A r\nu r(0)} + \Gamma^{r(0)}_{\nu A}\calL^{\mu A\nu r(0)} + \Gamma^{\nu(0)}_{\nu A}\calL^{\mu rAr(0)}
       + \Gamma^{r(0)}_{\nu A}\calL^{\mu r\nu A(0)} - \Gamma^{A(0)}_{A\nu}\calL^{\mu r\nu r(0)} \\
     &= \Gamma^{\mu(0)}_{\nu\rho}\calL^{\rho r\nu r(0)} + \Gamma^{r(0)}_{\nu r}\calL^{\mu r\nu r(0)} + \Gamma^{\nu(0)}_{\nu\rho}\calL^{\mu r\rho r(0)}
       + \Gamma^{r(0)}_{\nu r}\calL^{\mu r\nu r(0)} - 0 \\
     &= -\frac{3}{2}f'C_2 u^\mu \ . \label{eq:nabla_L_0}
 \end{split}
\end{align}
This establishes that $C_2 = 0$. The remaining (first) term in \eqref{eq:s_ambig} can now be written as:
\begin{align}
 s_{\mathrm{ambig.}}^{\mu(1)} ={}& -4\pi C_1\ell^\mu \left(\frac{1}{\kappa}\ell^\nu(\nabla_\nu\ell_r - \nabla_r\ell_\nu)\right)^{(1)} \ . \label{eq:s_ambig_1}
\end{align}
Just like in section \ref{sec:current:Einstein}, we can now force $\ell^\nu\nabla_r\ell_\nu = -\ell^\nu\nabla_\nu\ell_r$ by imposing the weak Killing condition \eqref{eq:killing_weak} at first order. We are left with:
\begin{align}
 s_{\mathrm{ambig.}}^{\mu(1)} ={}& -8\pi C_1\ell^\mu \left(\frac{1}{\kappa}\ell^\nu\nabla_\nu\ell_r\right)^{(1)}
   = -8\pi C_1\ell^\mu \left(\frac{\kappa}{\kappa}\right)^{(1)} = 0 \ .
\end{align}
We conclude that the Wald entropy current \eqref{eq:s} is well-defined at first order.

\subsection{Curvature-squared theories} \label{sec:current:R_squared}

In the previous subsection, we presented a formal argument for the vanishing of the transverse coefficient $C_2$ in \eqref{eq:L_0}. This was necessary to demonstrate that our entropy current is well-defined at first order. As a consistency check, we will now explicitly show that this is the case for curvature-squared Lagrangians. For a given solution, the condition $C_2 = 0$, i.e. $\calL^{\mu r\nu r(0)} \sim u^\mu u^\nu$, is linear in the Lagrangian. We can therefore check it for each Lagrangian term separately. The Einstein-Hilbert and cosmological terms satisfy the condition, as we've seen in section \ref{sec:current:Einstein}. The most general quadratic terms take the form:
\begin{align}
 \calL_1 = \sqrt{-g}R^{ABCD}R_{ABCD};\quad \calL_2 = \sqrt{-g}R^{AB}R_{AB};\quad \calL_3 = \sqrt{-g}R^2 \ .
\end{align}
Their derivatives with respect to the Riemann read:
\begin{align}
 \begin{split}
   \calL_1^{ABCD} &= 2\sqrt{-g}R^{ABCD} \\
   \calL_2^{ABCD} &= \frac{1}{2}\sqrt{-g}(R^{AC}g^{BD} - R^{AD}g^{BC} - R^{BC}g^{AD} + R^{BD}g^{AC}) \\
   \calL_3^{ABCD} &= \sqrt{-g}R(g^{AC}g^{BD} - g^{AD}g^{BC}) \ .
\end{split}
\end{align}
The relevant components for the question of ambiguities are $\calL^{\mu r\nu r}$:
\begin{align}
 \begin{split}
   \calL_1^{\mu r\nu r} &= 2\sqrt{-g}R^{\mu r\nu r} \\
   \calL_2^{\mu r\nu r} &= \frac{\sqrt{-g}}{2}(R^{\mu\nu}g^{rr} - R^{\mu r}g^{r\nu} - R^{r\nu}g^{\mu r} + R^{rr}g^{\mu\nu}) \\
   \calL_3^{\mu r\nu r} &= \sqrt{-g}R(g^{\mu\nu}g^{rr} - g^{\mu r}g^{r\nu}) \ .
 \end{split} \label{eq:L_123}
\end{align}
Let us now evaluate these quantities at zeroth order. We will need the volume density, the inverse metric and the Riemann tensor derived from the zeroth-order ansatz \eqref{eq:g_0}. The volume density is $\sqrt{-g^{(0)}} = \sqrt{-h}f^{3/2}$. 
The inverse metric on the horizon reads:
\begin{align}
   g^{(0)rr} = 0;\quad g^{(0)r\mu} = u^\mu;\quad g^{(0)\mu\nu} = \frac{1}{f} P^{\mu\nu} \ ,
\end{align}
while the independent nonvanishing components of $R^{(0)}_{ABCD}$ read:
\begin{align}
 R^{(0)}_{r\mu r\nu} = \frac{1}{2}\left(\left(\frac{f'^2}{2f} - f''\right)P_{\mu\nu} - k''u_\mu u_\nu\right);\quad R^{(0)}_{r\mu\nu\rho} = \frac{1}{2}f'k'P_{\mu[\nu}u_{\rho]} \ .
\end{align}
Plugging this into \eqref{eq:L_123}, we get:
\begin{align}
 \begin{split}
   \calL_1^{\mu r\nu r(0)} &= 2\sqrt{-h}f^{3/2}R^{\mu r\nu r(0)} = -\sqrt{-h}f^{3/2}k'' u^\mu u^\nu \\
   \calL_2^{\mu r\nu r(0)} &= \frac{1}{2}\sqrt{-h}f^{3/2}\left(0 - 2u^{(\mu}R^{\nu) r(0)} + \frac{1}{f}P^{\mu\nu} R^{rr(0)}\right) \\
     &= -\frac{1}{2}\sqrt{-h}f^{3/2}\left(k'' + \frac{3f'k'}{2f}\right)u^{(\mu} u^{\nu)} + 0 
     = -\frac{1}{2}\sqrt{-h}f^{3/2}\left(k'' + \frac{3f'k'}{2f}\right)u^\mu u^\nu \\
   \calL_3^{\mu r\nu r(0)} &= \sqrt{-h}f^{3/2}(0 - u^\mu u^\nu)R^{(0)} = -\sqrt{-h}f^{3/2}\left(k'' + \frac{3f'k'}{f}\right)u^\mu u^\nu \ ,
 \end{split}
\end{align}
and we find $\calL^{\mu r\nu r(0)} \sim u^\mu u^\nu$ as expected. 

\section{Einstein-Maxwell theory with gauge-gravitational Chern-Simons term} \label{sec:chern}

In section \ref{sec:current:condition}, we've seen that our proposed Wald entropy current is well-defined to first order in gradients for a class of higher-curvature theories. As the next step, we wish to examine the validity of the local Second Law 
$\del_\mu s^\mu \geq 0$ in a higher-curvature setup. It turns out that this task is relatively simple in a 5d theory of gravity coupled to gauge fields, with the higher-curvature effect coming from a mixed gauge-gravitational Chern-Simons term. The reason for the simplification is that in this setup, the higher-curvature term doesn't affect the ideal hydrodynamics, and only shows up in the next (viscous) order in the gradient expansion.

As explained in the Introduction, the hydrodynamics associated with this bulk theory is interesting in its own right. The gauge and gauge-gravitational Chern-Simons terms in the bulk action correspond to chiral and chiral-gravitational triangle anomalies in the QFT underlying the fluid.

We will find that our proposed entropy current for this theory is indeed locally increasing at leading order. Furthermore, we will find that it coincides with the hydrodynamic entropy current in the full fluid/gravity correspondence.

In this section, we specialize to a fluid living in a flat Minkowski metric, i.e. we set $h_{\mu\nu} = \eta_{\mu\nu}$. This does not involve a loss of generality. Indeed, it can be shown from index symmetries that the axial Chern-Simons terms cannot introduce into our calculations a contribution from the curvature tensor $\mathcal{R}_{\mu\nu\rho\sigma}$ associated with the metric $h_{\mu\nu}$. That such contributions do not come from other terms in the action was already established in \cite{Eling:2010hu}.

\subsection{Entropy current} \label{sec:chern:current}

In this subsection, we calculate our proposed entropy current \eqref{eq:s} for the aforementioned Einstein-Maxwell theory with Chern-Simons terms, up to first order in gradients.
We start with the usual Einstein-Maxwell Lagrangian, supplemented with pure gauge and mixed gauge-gravitational Chern-Simons terms:
\begin{align}
 \begin{split}
   \calL ={}& \frac{1}{16\pi}\left(\sqrt{-g}\left(R - 2\Lambda - F_{AB}F^{AB}\right) \vphantom{\frac{4}{3}} \right. \\
     &\left.+ 2\epsilon^{ABCDE}A^a_A\left(\frac{4}{3}\kappa_{abc}F^b_{BC}F^c_{DE} + \lambda_a R_{BCK}{}^L R_{DEL}{}^K \right) \right) \ .
 \end{split} \label{eq:L}
\end{align}
This is the same as eq. (2.6) in \cite{Landsteiner:2011iq}, except that $\Lambda$ is replaced with $-\Lambda$, and $F^a_{AB}$ is divided by $2$ to agree with the normalization conventions of \cite{Eling:2010hu}.

To find the Wald entropy current for this Lagrangian, we write down its derivative with respect to the Riemann tensor:
\begin{align}
 \begin{split}
   \calL^{ABCD} &\equiv \frac{\del\calL}{\del R_{ABCD}} = \calL^{ABCD}_{e.h.} + \calL^{ABCD}_{c.s.} \\
     &= \frac{1}{16\pi}\left(\sqrt{-g}g^{A[C}g^{D]B} - 2\lambda_a A^a_K\left(\epsilon^{KLMAB}R_{LM}{}^{CD} + \epsilon^{KLMCD}R_{LM}{}^{AB}\right) \right) \ .
 \end{split}
\end{align}

The $\lambda_a$ term satisfies the $A\leftrightarrow B$ and $C\leftrightarrow D$ antisymmetries and the $(AB)\leftrightarrow (CD)$ symmetry of the Riemann tensor, but not the Bianchi symmetry. The latter will prove immaterial to our definition of the entropy current. We can enforce the Bianchi symmetry by redefining $\calL^{ABCD} \rightarrow \calL^{ABCD}-\calL^{[ABCD]}$. However, we will see in a moment that the extra $\calL^{[ABCD]}$ term does not contribute to the definition of the entropy current.

Our entropy current was defined in section \ref{sec:current:definition} to be:
\begin{align}\label{eq:def_entropy_again}
  s^\mu = \frac{2\pi}{\kappa}\left(-2\calL^{\mu r\nu r}(\nabla_\nu\ell_r - \nabla_r\ell_\nu) + 4\ell_r\nabla_\nu\calL^{\mu r\nu r}\right) \ .
\end{align}
Since this definition is linear in $\calL^{ABCD}$, it is enough to show that an antisymmetrized $\calL^{[ABCD]}$ term does not, by itself, contribute to the entropy current. This is so, because both $L^{\mu r\nu r}$ and $\nabla_\nu L^{\mu r\nu r}$ vanish when antisymmetrized over the two identical $r$ indices.

Due to its linearity in $\calL^{\mu r\nu r}$, we can separate the entropy current into two parts:
\begin{align}
  s^\mu = s^\mu_{e.h.} + s^\mu_{c.s.} \ ,
\end{align}
where $s^\mu_{e.h.}$ is the part of the entropy current coming from the Einstein-Hilbert part of the Lagrangian, and $s^\mu_{c.s.}$ is the part arising from the Chern-Simons terms. The Einstein-Hilbert part $s^\mu_{e.h.}$ is given in \eqref{eq:f_s_current}.
 In a flat metric $h_{\mu\nu} = \eta_{\mu\nu}$, it simplifies into:
\begin{align}
 s^\mu_{e.h.} = su^\mu \ .
\end{align}
In general, we define the entropy density $s$ as the norm of the entropy current under the flat metric. As we will see, the Chern-Simons contribution $s^\mu_{s.c.}$ is first-order in gradients and transverse to $u^\mu$. Therefore, it doesn't affect the entropy density at first order.

We now turn to the higher-curvature contribution $s^\mu_{c.s.}$. Using the zeroth-order hydrodynamic ansatz \eqref{eq:g_0} and the Christoffel symbols found in \eqref{eq:Christoffel_0}, one can find by direct calculation that $R^{(0)}_{\mu\nu}{}^{\rho r} = 0$, and therefore $\calL_{c.s.}^{\mu r\nu r(0)} = 0$. From the discussion in section \ref{sec:current:condition}, it follows that there is no zeroth-order contribution $s^{\mu(0)}_{c.s.}$ to the entropy current from the Chern-Simons term. Our task, therefore, is to calculate the first-order contribution $s^{\mu(1)}_{c.s.}$.

Before we get started, we must explicitly introduce the gauge potential $A_A^a$ into our hydrodynamic ansatz.
By a choice of gauge, we can force $A^{a}_r = 0$ everywhere at all orders. With this choice, the zeroth-order gauge potential in the vicinity of the horizon reads:
\begin{align}
 A^{a(0)}_\mu = a^a(r,x^\mu)u_\mu \ . \label{eq:A_0}
\end{align}
The horizon value $a^a(0,x^\mu)=-\mu^a(x^\mu)$ encodes the horizon's chemical potential, while the radial derivative $a_a'(0) = \pi n_a/s$ is proportional to the charge density $n_a$ (cf. \cite{Eling:2010hu}).

Alongside the metric corrections $g^{(1)}_{\mu\nu}$ which were discussed in section \ref{sec:pre:hydro}, we must also consider corrections $A^{a(1)}_\mu$ to the gauge potential \eqref{eq:A_0}. It turns out, however, that neither type of
correction affects the entropy current $s^\mu_{c.s.}$. To evaluate it, we will need the Riemann components:
\begin{align}
 \begin{split}
   R^{(1)}_{\mu\nu}{}^{\rho r} = \frac{1}{2}u^\rho \del_\mu(k'u_\nu) + \frac{k'}{4}u_\nu\left[2P^{\rho\sigma} \del_{(\sigma}u_{\mu)} +\frac{1}{f}P_\mu^\rho u^\delta \del_\delta f\right] - (\mu\leftrightarrow\nu) \ .
 \end{split}
\end{align}
From these we calculate:
\begin{align}
 \begin{split}
    \calL_{c.s.}^{\mu r\nu r(1)}&
     = -\frac{\lambda_a}{8\pi} A^{a(0)}_K\left(\epsilon^{KLM\mu r}R^{(1)}_{LM}{}^{\nu r} + \epsilon^{KLM\nu r}R^{(1)}_{LM}{}^{\mu r} \right) \\
     &=-\frac{k'}{4\pi}\lambda_a a^a\epsilon^{\kappa\rho\sigma(\mu} u^{\nu)} u_\kappa\del_\rho u_\sigma
   = \frac{k'}{2\pi}\lambda_a a^a\omega^{(\mu}u^{\nu)} \ ,
 \end{split} \label{eq:L_mu_r_nu}
\end{align}
where the vorticity field $\omega^\mu$ is defined as $\omega^\mu = \frac{1}{2}\epsilon^{\mu\nu\rho\sigma} u_\nu \del_\rho u_\sigma$.

We can now calculate the $\calL_{c.s.}^{ \mu r\nu r (1)}(\nabla_\nu\ell_r - \nabla_r\ell_\nu){}^{(0)}$ piece of $s^\mu_{c.s.}$ (see \eqref{eq:def_entropy_again}). At zeroth order, the horizon normal $\ell^A = (0,u^\mu)$ satisfies the full Killing condition $\nabla_A\ell_B = -\nabla_B\ell_A$.
Using this, we get:
\begin{align}
 \begin{split}
 \calL_{c.s.}^{ \mu r\nu r (1)}(\nabla_\nu\ell_r - \nabla_r\ell_\nu){}^{(0)}=
 2\calL_{c.s.}^{ \mu r\nu r (1)}(\nabla_\nu\ell_r){}^{(0)}=
 -2\calL_{c.s.}^{ \mu r\nu r (1)} \Gamma_{\nu r}^{r(0)}
 =-\frac{k'^2}{4\pi}\lambda_a a^a\omega^{\mu}\label{eq:s_piece_1}
 \ ,
 \end{split}
\end{align}
where we plugged in the Christoffel symbols from \eqref{eq:Christoffel_0}.

We now turn to the second piece of $s^\mu_{c.s.}$, which is proportional to:
\begin{align}
\begin{split}
\nabla_\nu\calL_{c.s.}^{\mu r\nu r} = &
     \del_\nu\calL_{c.s.}^{\mu r\nu r} + \Gamma^\mu_{\nu \rho}\calL_{c.s.}^{\rho r\nu r} +
     \Gamma^{r}_{\nu A}\calL_{c.s.}^{\mu A\nu r} +
      \Gamma^{\nu}_{\nu \rho}\calL_{c.s.}^{\mu r \rho r} + \Gamma^{r}_{\nu A}\calL_{c.s.}^{\mu r\nu A} - \Gamma^{A}_{A\nu}\calL_{c.s.}^{\mu r\nu r} \\
     = & \del_\nu\calL_{c.s.}^{\mu r\nu r} + \Gamma^\mu_{\nu \rho}\calL_{c.s.}^{\rho r\nu r} +
     \Gamma^{r}_{\nu \alpha}\calL_{c.s.}^{\mu \alpha \nu r}
      + \Gamma^{r}_{\nu r}\calL_{c.s.}^{\mu r\nu r} +\Gamma^{r}_{\nu \alpha}\calL_{c.s.}^{\mu r\nu \alpha}\ .
\end{split}\label{eq:nabla_L}
\end{align}
As expected, at zeroth order this expression vanishes, since both $\Gamma^{r(0)}_{\mu\nu}$ and $\calL_{c.s.}^{\mu r\nu r(0)}$ vanish.
 To obtain the first-order expression, we need the quantity:
\begin{align}
 \begin{split}
   \calL_{c.s.}^{\mu\nu\rho r(0)} &=
    -\frac{\lambda_a}{8\pi} A^{a(0)}_K\left(\epsilon^{KLM\mu\nu}R^{(0)}_{LM}{}^{\rho r} + \epsilon^{KLM\rho r}R^{(0)}_{LM}{}^{\mu\nu} \right) \\
   &= -\frac{\lambda_a a^a}{8\pi}u_\kappa \left(-2\epsilon^{\kappa\sigma\mu\nu}R^{(0)}_{r\sigma}{}^{\rho r}
      + \epsilon^{\kappa\lambda\sigma\rho}R^{(0)}_{\lambda\sigma}{}^{\mu\nu} \right)=
      -\frac{f'k'}{16\pi f}\lambda_a a^a\epsilon^{\kappa\rho\mu\nu}u_\kappa \ ,
 \end{split} \label{eq:L_mu_nu_rho r}
\end{align}
where we used the Riemann components:
\begin{align}
\begin{split}
&R_{\mu\nu}^{(0)}{}^{\rho\sigma}=
\frac{f'k'}{4f}u_\mu(u^\sigma P_\nu^\rho-u^\rho P_\nu^\sigma)-(\mu \leftrightarrow \nu)\\
& R^{(0)}_{r\mu}{}^{\nu r} = \frac{1}{2}k'' u_\mu u^\nu - \frac{f'k'}{4f}P_\mu^\nu \ .
\end{split}
\end{align}
We note that $\calL_{c.s.}^{\mu\nu\rho r(0)}$ in \eqref{eq:L_mu_nu_rho r} is antisymmetric in all its Greek indices, so the corresponding terms in \eqref{eq:nabla_L} vanishes. We are left with:
\begin{align}
\begin{split}
 \left(\nabla_\nu\calL_{c.s.}^{\mu r\nu r} \right)^{(1)}= &
 \Gamma^{\mu(0)}_{\nu \rho}\calL_{c.s.}^{\rho r\nu r(1)}+ \Gamma^{r(0)}_{\nu r}\calL_{c.s.}^{\mu r\nu r(1)} =
      0-\frac{k'^2}{4\pi} u_\nu \lambda_a a^a\omega^{(\mu}u^{\nu)} = \frac{k'^2}{8\pi}\lambda_a a^a\omega^\mu \ .
      \label{eq:s_piece_2}
\end{split}
\end{align}

Plugging \eqref{eq:s_piece_1} and \eqref{eq:s_piece_2} into \eqref{eq:def_entropy_again}, we finally get:
\begin{align}
 s_{c.s.}^{\mu(1)} &=
  \frac{2\pi}{\kappa}\left(\frac{k'^2}{2\pi}\lambda_a a^a\omega^{\mu}+
  \frac{k'^2}{2\pi}\lambda_a a^a\omega^\mu\right) =
  -16\pi T\lambda_a \mu^a\omega^{\mu} \ ,
\end{align}
where we used eq. \eqref{eq:temperature_def} to relate $k'$ to the temperature at zeroth order.
Thus, the final result for the Wald entropy current reads:
\begin{align}\label{eq:entropy_found}
 s^\mu &= s_{e.h.}^\mu + s_{c.s.}^\mu = su^\mu - 16\pi T\lambda_a \mu^a\omega^{\mu} \ .
\end{align}

\subsection{The entropy production rate}\label{subsec:entropy_production}

As we recall from the beginning of the section, our goal is to demonstrate that the proposed Wald entropy current for the theory \eqref{eq:L}, which we evaluated as \eqref{eq:entropy_found}, has a non-negative divergence.

As discussed in section \ref{sec:pre:BH}, the divergence of the area current $J^\mu_{Area}=\nobreak\sqrt{-g}\ell^\mu = \nobreak 4s u^\mu$
is governed by the focusing equation (the twice-projected component of the Einstein equation $E_{AB}\ell^A\ell^B$). In section \ref{sec:pre:hydro}, we used the detailed form of the focusing equation \eqref{eq:focusing} in Einstein gravity to prove that the local divergence of the entropy current \eqref{eq:Bekenstein_current} is non-negative at leading nontrivial order in the hydrodynamic expansion.

When we add gauge-gravitational Chern-Simons terms to the Lagrangian, the Einstein equations get an extra contribution. This leads to new contributions to the focusing equation, which will allow us to prove that the entropy current \eqref{eq:entropy_found} satisfies a local increase law.

We start from the Einstein equation derived by varying \eqref{eq:L} with respect to the metric:
\begin{align}
 \begin{split}
   \sqrt{-g}\left(G_A^B + \Lambda\delta_A^B\right) ={}& \sqrt{-g}\left(2F^a_{AC}F_a^{BC} - \frac{1}{2}F^2\delta_A^B\right) \\
   &+ 2\lambda_a\left(g_{AC}g^{BD} + \delta^B_C\delta^D_A\right)\epsilon^{CKLMN}\nabla_E\left(F^a_{KL}R_{MND}{}^E\right) \ .
 \end{split} \label{eq:einstein}
\end{align}

This is the same as eq. (2.16) in \cite{Landsteiner:2011iq}, except that (as mentioned before) $F^a_{AB}$ is divided by 2, and $\Lambda$ is replaced with $-\Lambda$.

When projected twice along $\ell^A$, the first row of \eqref{eq:einstein} gives us the focusing equation \eqref{eq:focusing} times $\sqrt{-g}$, with the matter stress tensor $\mathcal{T}_{AB} = (1/4\pi)\left(F^a_{AC}F_a^{BC} - \frac{1}{4}F^2\delta_A^B\right)$ (which satisfies the null energy condition).

In appendix \ref{app:extra_focusing}, we calculate the contribution of the second line of \eqref{eq:einstein} to the focusing equation. This turns out to vanish at zeroth and first order in derivatives. The second-order contribution reads:
\begin{align}
 \begin{split}
     \mbox{new term}^{(2)} {} = {} & 2\lambda_a \ell^A \ell_B \left(g_{AC}g^{BD} + \delta^B_C\delta^D_A\right)\epsilon^{CKLMN}\nabla_E\left(F^a_{KL}R_{MND}{}^E\right)\\
      = {}&  2\lambda_a u^\alpha \left(g_{\alpha C} \delta^{r}_{D} +
      \delta^r_C g_{D \alpha}\right)\epsilon^{CKLMN}\nabla_E\left(F^a_{KL}R_{MN}{}^{DE}\right)\\
      ={} &   64\lambda_a \pi^2 T\epsilon^{\kappa \lambda \mu \nu }\del_\kappa(\mu^a u_\lambda)\del_\mu(Tu_\nu)
     =128 \lambda_a\pi^2 T \del_\nu(\mu^a T \omega^\nu ) \ .
 \end{split}
\end{align}

Thus, at first order the focusing equation \eqref{eq:focusing_1} remains unchanged, and we still have
$\theta^{(1)}=0$. Using this, the second-order focusing equation reads:
\begin{align}
 \kappa^{(0)}\theta^{(2)} = (\sigma_{(H)}^{(1)})^2 + 8\pi\mathcal{T}^{(2)}_{AB}\ell^A\ell^B +
 \frac{1}{\sqrt{-g}} 128\lambda_a \pi^2 T \del_\nu(\mu^a T \omega^\nu) \ ,
\end{align}
where $\theta = \del_\mu J^\mu_{Area}/\sqrt{-g} = 4\del_\mu(s u^\mu)/\sqrt{-g}$.
 After a short rearrangement, the focusing equation becomes:
\begin{align}
  \del_\mu \left(s u^\mu - 16\pi\lambda_a\mu^a T \omega^\mu \right) = \frac{\sqrt{-g}}{8\pi T} \left[ (\sigma_{(H)}^{(1)})^2 + 8\pi\mathcal{T}^{(2)}_{AB}\ell^A\ell^B \right] \geq 0 \ .
\end{align}
We have thus demonstrated that our proposed entropy current \eqref{eq:entropy_found} is indeed locally increasing.

\section{Anomalous hydrodynamics from horizon dynamics} \label{sec:brane}

\subsection{Introduction}

We've seen that a horizon projection of the Einstein equations in a hydrodynamic limit of the higher-curvature theory \eqref{eq:L} encodes the local increase of the Wald entropy current \eqref{eq:entropy_found}. In this section, we will demonstrate that a broader selection of the horizon-projected field equations encodes the full equations of charged (anomalous) hydrodynamics. Our proposed Wald entropy current then corresponds precisely to the entropy current as defined in the hydrodynamics.

The present result extends our previous calculations \cite{Eling:2010hu} without the higher-curvature Chern-Simons term. As in \cite{Eling:2010hu}, the exact correspondence between the horizon field equations and hydro equations only holds under certain constraints on the corrections $g^{(1)}_{\mu\nu}, A^{(1)}_\mu$ to the fluid ansatz. In AdS/CFT, these constraints are enforced by the boundary conditions at infinity. Apart from assuming the constraints, our horizon-based method does not require detailed knowledge of the corrections, because the \emph{same} hydrodynamics emerges for all valid choices of $g^{(1)}_{\mu\nu}$ and $A^{(1)}_\mu$.
In particular, in an AdS/CFT context, the hydrodynamics which we find from the horizon analysis will be the same as the one on the AdS boundary.

The discussion thus far has been driven by the concept of a Wald entropy current. However, the hydrodynamics arising from the theory \eqref{eq:L} is of interest in a broader context. The $\lambda_a$ term in the bulk action generates a peculiar vortical conductivity term $J^\mu \sim T^2\omega^\mu$, generating currents along the vortices in the 4d fluid.
It is conjectured \cite{Landsteiner:2011cp,Landsteiner:2011iq} that such terms are directly related to mixed chiral-gravitational anomalies in the QFT underlying the fluid. In particular, the authors of \cite{Landsteiner:2011iq} have established this relation in a specific AdS/CFT setup. There, the 4d hydrodynamics is embedded in a conformal field theory on the 4d boundary of the AdS space. The bulk Chern-Simons coefficients $\kappa_{abc},\lambda_a$ correspond respectively to chiral and mixed chiral-gravitational anomalies in the CFT. The precise relation is given by eqs. (2.9)-(2.14) in \cite{Landsteiner:2011iq} (for the conventions in the definition of the anomaly coefficients, see eqs. (30)-(32) in \cite{Landsteiner:2011cp}):
\begin{align}
 \kappa_{abc} &= -\frac{1}{4\pi}\left[ \tr(T_a\{T_b,T_c\})_R - \tr(T_a\{T_b,T_c\})_L \right] \\
 \lambda_{a} &= -\frac{1}{48\pi}\left[ \tr(T_a)_R - \tr(T_a)_L \right] \ ,
\end{align}
where $T_a$ are the generators of the global symmetry group under which the chiral fermions in the CFT transform. The authors of \cite{Landsteiner:2011iq} then derived the relation between the $\lambda_a$ Chern-Simons term and the coefficient of the vortical conductivity term, thus relating the latter to the chiral/gravitational anomaly. Our calculation below generalizes the results of \cite{Landsteiner:2011iq} on the relation between the bulk Chern-Simons term and the fluid's vortical conductivity. We derive the result for a more general fluid/gravity ansatz, which allows in particular for an arbitrary (not necessarily conformal) equation of state.

\subsection{Hydrodynamics from the horizon field equations} \label{sec:full_hydro}
We will now detail the relation between the horizon field equations and hydrodynamics. In particular, we wish to recast the field equations into the form of hydrodynamic conservation laws. From these one can later extract an entropy current. We begin by writing down the Einstein and Maxwell field equations as derived from the Lagrangian \eqref{eq:L}:
\begin{align}
 \begin{split}
   \sqrt{-g}\left(G_A^B + \Lambda\delta_A^B\right) ={}& \sqrt{-g}\left(2F^a_{AC}F_a^{BC} - \frac{1}{2}F^2\delta_A^B\right) \\
     &+ 2\lambda_a\left(g_{AC}g^{BD} + \delta^B_C\delta^D_A\right)\epsilon^{CKLMN}\nabla_E\left(F^a_{KL}R_{MND}{}^E\right)
 \end{split} \label{eq:einstein2} \\
 \del_B\left(\sqrt{-g}F_a^{AB}\right) ={}& \epsilon^{AKLMN}\left(2\kappa_{abc} F^b_{KL} F^c_{MN} + \frac{\lambda_a}{2} R_{KLC}{}^D R_{MND}{}^C \right) \ . \label{eq:maxwell}
\end{align}
These correspond to eqs. (2.16)-(2.17) in \cite{Landsteiner:2011iq}.

The horizon projection of these equations without the $\lambda_a$ terms was studied previously in \cite{Eling:2010hu}.
We briefly summarize the results.
Writing eqs. \eqref{eq:einstein2}-\eqref{eq:maxwell} symbolically as $E_A^B$ and $M_a^A$, the relevant components for the hydrodynamics are $E_\mu^r$ and $M_a^r$, evaluated on the horizon. Then, to second order in small gradients, certain linear combinations of these equations take the form of hydrodynamic conservation laws for a stress tensor $T^{\mu\nu}$ and a set of charge currents $J_a^\mu$:
\begin{align}
 \begin{split}
   -\frac{1}{8\pi}E_\mu^r + \frac{1}{4\pi}\mu^a u_\mu M_a^r \quad &\Rightarrow \quad \del_\nu T^\nu_\mu = 0 \\
   \frac{1}{4\pi}M_a^r \quad &\Rightarrow \quad \del_\mu J_a^\mu = 0 \ .
 \end{split} \label{eq:equivalence}
\end{align}
More precisely, the field equations take the form (section V in \cite{Eling:2010hu}):
\begin{align}
  E_\mu^r:\quad -8\pi(\del_\nu T_\mu^\nu - \mu^a u_\mu\del_\nu J^\nu_a) &= \Delta_\mu
    + 8su^\nu\del_{[\mu}c^{(1)}_{\nu]} + 16\pi n_a u^\nu\del_{[\mu}A^{a(1)}_{\nu]} \label{eq:GC_with_corrections} \\
  M_a^r:\quad 4\pi\del_\mu J_a^\mu &= 0 \ . \label{eq:Max_with_corrections}
\end{align}

The equations depend on the corrections to the fluid/gravity ansatz \eqref{eq:g_0},\eqref{eq:A_0} through the gauge field correction $A^{a(1)}_\mu$ and the extrinsic curvature quantity $c^{(1)}_\mu \equiv -\Gamma^{\nu(1)}_{\mu\rho}u_\nu u^\rho = - \left(1/2\right) u^\nu(\del_\nu u_\mu + g'^{(1)}_{\mu\nu})$, which appear on the RHS of \eqref{eq:GC_with_corrections}. The $\Delta_\mu$ in \eqref{eq:GC_with_corrections} stands for additional terms, which do not fit into the conservation-law template on the LHS, but which can be canceled by an appropriate choice of the corrections $g'^{(1)}_{\mu\nu}u^\nu$ and $A^{a(1)}_\mu$ (or equivalently, $c^{(1)}_\mu$ and $A^{a(1)}_\mu$). With such a choice, we recover the conservation laws in \eqref{eq:equivalence}.

As explained in \cite{Eling:2010hu} (see the text surrounding eqs. (V.60)-(V.61)), the correction terms cannot themselves produce a contribution to the divergence $\del_\nu T_\mu^\nu$. Thus, despite the remaining freedom in choosing the corrections, there is no ambiguity in the stress tensor $T^\nu_\mu$ and the charge currents $J_a^\mu$, which define the hydrodynamics.

As calculated in \cite{Eling:2010hu} (eqs. (V.21),(V.18),(III.20),(V.63)), the stress tensor and charge currents in \eqref{eq:equivalence}-\eqref{eq:Max_with_corrections} are given by the constitutive relations:
\begin{align}
  T^{\mu\nu} &= \epsilon u^\mu u^\nu + p P^{\mu\nu} - \frac{s}{2\pi}\pi^{\mu\nu} - 2T\mu^a\tilde\sigma_{ab}u^{(\mu} P^{\nu)\rho}\del_\rho\frac{\mu^b}{T}
    - \frac{16}{3\pi}\kappa_{abc}\mu^a\mu^b\mu^c u^{(\mu}\omega^{\nu)} \ , \label{eq:T_constit} \\
  J_a^\mu &= n_a u^\mu - T\tilde\sigma_{ab}P^{\mu\nu}\del_\nu\frac{\mu^b}{T}
    - \frac{4}{\pi}\kappa_{abc}\mu^b\mu^c\omega^\mu \ , \label{eq:J_constit} \\
  \tilde\sigma_{ab} &= \frac{s^{1/3}}{2^{4/3}\pi}\left(\delta_{ab} - \frac{\mu_a n_b}{\epsilon + p}\right) \label{eq:tilde_sigma_ab}\ .
\end{align}
In \eqref{eq:T_constit} and \eqref{eq:tilde_sigma_ab}, the energy density $\epsilon$ and the pressure $p$ are derived from the potentials and densities $T,\mu_a,s,n^a$ on the horizon via thermodynamic identities, after assuming an equation of state.
$\pi^{\mu\nu}$ is the shear tensor of the fluid's velocity, $\pi^{\mu\nu} \equiv P^{\mu\rho}P^{\nu\sigma}\del_{(\rho}u_{\sigma)} - (1/3)P^{\mu\nu}\del_\rho u^\rho$. $\tilde\sigma_{ab}$ is a conductivity matrix.

These equations define charged relativistic hydrodynamics, with vanishing bulk viscosity and special values for the shear viscosity and for the charge conductivity. The $\kappa_{abc}$ term corresponds to chiral anomalies in the underlying QFT. The constitutive relations \eqref{eq:T_constit}-\eqref{eq:J_constit} are written in a slightly nonstandard form, since our velocity variable $u^\mu$ is neither the energy velocity nor the charge velocity of the fluid. In fact, the horizon normal $u^\mu$ corresponds to the fluid's entropy velocity. 
This will no longer be the case once the higher-curvature $\lambda_a$ terms are included.

Our present goal is to find how the above equivalence to hydrodynamics is modified in the presence of the gauge-gravitational Chern-Simons term $\lambda_a$ in the field equations \eqref{eq:einstein2}-\eqref{eq:maxwell}.

The $\lambda_a$ terms on the RHS of the field equations are evaluated in appendices \ref{app:Maxwell} and \ref{app:GC}, and read:
\begin{align}
\begin{split}\label{eq:GC_to_check1}
 \delta E^r_\mu &= -64 \pi^2 \lambda_a Tu_\mu\epsilon^{\nu\rho\sigma\lambda}\del_\nu(\mu^a u_\rho)\del_\sigma(Tu_\lambda)
   - 16\lambda_a s u^\nu \del_{[\mu}\left(\frac{Q^a}{s} \omega_{\nu]} \right)
\end{split}\\
\begin{split}\label{eq:maxwell_to_check1}
 \delta M^r_a &= 16\pi^2 \lambda_a \epsilon^{\mu\nu\rho\sigma}\del_\mu(Tu_\nu)\del_\rho(Tu_\sigma) \ ,
\end{split}
\end{align}
where the function $Q^a \equiv f^2 (k'a^a/f)'$ in the second term in \eqref{eq:GC_to_check1} is unfortunately not expressible in terms of the thermal parameters $T,\mu_a,s,n^a$.
However, we can see from \eqref{eq:GC_with_corrections} that this function can be tuned arbitrarily by further adjusting the first-order corrections $c^{(1)}_\mu$
(the $A^{(1)}_\mu$ term can also fulfill this purpose).
Thus, by an appropriate choice of corrections,
we can turn eqs. \eqref{eq:GC_to_check1}-\eqref{eq:maxwell_to_check1} into:
\begin{align}
\begin{split}\label{eq:GC_to_check2}
 \delta E^r_\mu &= -64 \pi^2 \lambda_a Tu_\mu\epsilon^{\nu\rho\sigma\lambda}\del_\nu(\mu^a u_\rho)\del_\sigma(Tu_\lambda)
   + 256\pi^2\lambda_a s u^\nu \del_{[\mu}\left(\frac{\mu^a T^2}{s} \omega_{\nu]} \right)
\end{split}\\
\begin{split}\label{eq:maxwell_to_check2}
 \delta M^r_a &= 16\pi^2 \lambda_a \epsilon^{\mu\nu\rho\sigma}\del_\mu(Tu_\nu)\del_\rho(Tu_\sigma) \ .
\end{split}
\end{align}

With this choice of the second term in \eqref{eq:GC_to_check2}, the equivalence \eqref{eq:equivalence} to hydrodynamic conservation laws is maintained. This is demonstrated in appendix \ref{app:new_constit}. The constitutive relations \eqref{eq:T_constit}-\eqref{eq:J_constit} for the stress tensor and charge currents acquire the following extra contributions:
\begin{align}\label{eq:new_constit}
\begin{split}
 \delta T^{\mu\nu} &= -32\pi\lambda_a\mu^a T^2 u^{(\mu}\omega^{\nu)} \\
 \delta J_a^\mu &= -8\pi\lambda_a T^2\omega^\mu.
\end{split}
\end{align}

For comparison with the literature, it is useful to rewrite the new constitutive relations in terms of an energy velocity $v^\mu$, instead of the horizon normal $u^\mu$. The result is:
\begin{align}
\begin{split}\label{eq:T_constit_final}
  T^{\mu\nu} ={}& (\epsilon + p)v^\mu v^\nu + p \eta^{\mu\nu} - \frac{s}{2\pi}\pi^{\mu\nu} \ ,
\end{split}\\
\begin{split} \label{eq:J_constit_final}
  J_a^\mu ={}& n_a v^\mu - T\sigma_{ab}P^{\mu\nu}\del_\nu\frac{\mu^b}{T}
  \\& - \left(\frac{4}{\pi}\kappa_{abc}\mu^b\mu^c + 8\pi\lambda_a T^2
   - \frac{8n_a}{\epsilon + p}\left(\frac{1}{3\pi}\kappa_{bcd}\mu^b\mu^c\mu^d + 2\pi\lambda_b\mu^b T^2\right)\right)\omega^\mu \ ,
\end{split}\\
\begin{split}
  \sigma_{ab} ={}& \left(\delta_a^c - \frac{n_a\mu^c}{\epsilon + p} \right) \tilde\sigma_{cb}
   = \frac{s^{1/3}}{2^{4/3}\pi}\left(\delta_{ab} - \frac{2\mu_{(a} n_{b)}}{\epsilon + p}
    + \frac{\mu_c\mu^c n_a n_b}{(\epsilon + p)^2}\right) \ ,
\end{split}
\end{align}
where the (first-order) difference between the velocities $u^\mu$ and $v^\mu$ is given by:
\begin{align}
 v^\mu &= u^\mu - \frac{\mu_a}{s}\sigma^{ab}P^{\mu\nu}\del_\nu\frac{\mu_b}{T}
    -\frac{8}{\epsilon + p}\omega^\mu\left(\frac{1}{3\pi}\kappa_{abc}\mu^a\mu^b\mu^c + 2\pi\lambda_a\mu^a T^2 \right) \ . \label{eq:v_u}
\end{align}

These constitutive relations are of the general form given in \cite{Neiman:2011} for the hydrodynamics of a 4d field theory with chiral anomalies (as derived from entropic constraints).
Thus, we have demonstrated a full correspondence between our gravitational model and a 4d fluid.
The new $\lambda_a$-proportional terms have the same form as the $\beta_a$ terms in \cite{Neiman:2011}, which have been related in \cite{Landsteiner:2011iq,Landsteiner:2011cp} to chiral/gravitational anomalies.

\subsection{Entropy current from the hydrodynamics}

Given a set of hydrodynamic conservation laws, one should be able to deduce an entropy current with positive divergence. For constitutive relations of the form \eqref{eq:T_constit_final}-\eqref{eq:J_constit_final}, in particular with the $\lambda_a$ terms, this has been done in \cite{Neiman:2011}. In terms of the energy velocity $v^\mu$, the result reads:
\begin{align}
  s^\mu = sv^\mu + \mu_a\sigma^{ab}P^{\mu\nu}\del_\nu\frac{\mu_b}{T}
    +\frac{8}{\epsilon+p} \omega^\mu\left(\frac{s}{3\pi }\kappa_{abc}\mu^a\mu^b\mu^c - 2\pi\mu_a n^a \lambda_b \mu^b T \right) \ . \label{eq:s_v}
\end{align}
Using eq. \eqref{eq:v_u} to translate back from the energy velocity $v^\mu$ to the horizon normal $u^\mu$, \eqref{eq:s_v} simplifies into:
\begin{align}
  s^\mu = su^\mu - 16\pi T\lambda_a \mu^a\omega^{\mu} \ .
\end{align}
This is in precise agreement with eq. \eqref{eq:entropy_found}, which we derived directly from our definition of a generalized Wald entropy current.

In fact, given the non-negative divergence of the Wald entropy current and the equivalence between the gravitational and hydrodynamic equations, it necessarily follows that the two entropy currents match. The reason is that, to first order in gradients, a fluid of the type discussed in \cite{Neiman:2011} allows just \emph{one} non-decreasing entropy current to be constructed from the hydrodynamic variables.

\section{Summary and discussion}
The local entropy current in Einstein gravity is a vector density directed along the horizon's generating light rays, whose flux through a spatial slice of the horizon is $1/4$ of the area of the slice. This entropy current maps via the fluid/gravity correspondence to the hydrodynamic entropy current of a field theory.
Working in the framework of the fluid/gravity correspondence, we proposed a definition for a local horizon entropy current in higher-curvature gravitational theories.
This current is a vector density on the horizon hypersurface, such that for stationary solutions its flux equals the total Wald entropy.
We demonstrated that our entropy current is well-defined to first order in fluid gradients, for actions with an algebraic dependence on the curvature.
It is of interest to carry the analysis further and study the properties of the current at higher orders in fluid gradients.

As an application, we considered the correspondence between 5d Einstein-Maxwell theory with gauge and gauge-gravitational Chern-Simons terms and 4d anomalous field theory hydrodynamics, which has attracted much interest recently.
The chiral and mixed chiral-gravitational anomalies in the field theory exhibit new non-dissipative transport coefficients, which are responsible for the generation of current along magnetic fields and vortices in the fluid.
In \cite{Neiman:2011}, the coefficient of the vortical term $T^2\omega^\mu$ in the current came about as a constant of integration, and was left arbitrary.
It has since been calculated in two regimes: in the free-fermions limit \cite{Landsteiner:2011cp}, and in the strong-coupling supergravity limit \cite{Landsteiner:2011iq}.
The vortical conductivity was found to be proportional to the chiral/gravitational anomaly, with the same proportionality constant in the two regimes.

In this work, we carried out the analysis of the horizon dynamics along the lines of \cite{Eling:2010hu}, and established fully the fluid/gravity duality between the anomalous quantum field theory hydrodynamics and the dynamics of an Einstein-Maxwell black brane with a gauge-gravitational Chern-Simons term.
In particular, the entropy current as defined in the field theory hydrodynamics coincides with our proposed Wald entropy current.
The results provide further support to the relationship between the vortical conductivity and the mixed chiral/gravitational triangle anomaly.
In particular, our analysis of the horizon dynamics was not restricted to conformal fluids, thus supporting a rather general relation between the vortical conductivity and the chiral/gravitational anomaly. Still, it remains to be proven that this holds at arbitrary coupling.

\section*{Acknowledgements}

This work is supported in part by the Israeli Science Foundation center of excellence and by the German-Israeli Foundation (GIF).

\appendix

\section{Chern-Simons contribution to the focusing equation}\label{app:extra_focusing}

In this appendix we calculate the contribution to the focusing equation from the mixed gauge-gravitational Chern-Simons term.
This contribution comes from the last term on the RHS of \eqref{eq:einstein}, twice contracted with the horizon normal:
\begin{align}
 \begin{split}
 &2\lambda_a \ell^A \ell_B \left(g_{AC}g^{BD} + \delta^B_C\delta^D_A\right)\epsilon^{CKLMN}\nabla_E\left(F^a_{KL}R_{MND}{}^E\right)=\\
 &2\lambda_a u^\alpha \left(g_{\alpha C}\delta^{r}_{D} +
 \delta^r_C g_{D\alpha}\right)\epsilon^{CKLMN}\nabla_E\left(F^a_{KL}R_{MN}{}^{DE}\right) \ .
 \end{split}
\end{align}

Before contracting with $u^\alpha$, let us write down the full expression for
\\$\left(g_{\alpha C}\delta^{r}_{D} +
\delta^r_C g_{D\alpha}\right)\epsilon^{CKLMN}\nabla_E\left(F^a_{KL}R_{MN}{}^{DE}\right)$. It will be needed in appendix
\ref{app:GC} as well.
\begin{align}
\begin{split}\label{eq:GC_full}
 & \left(g_{\alpha C}\delta^{r}_{D} + \delta^r_C  g_{D\alpha}\right) \epsilon^{CKLMN}\nabla_E\left(F^a_{KL}R_{MN}{}^{DE}\right)  \\
   & = \left(g_{\alpha C}\delta^{r}_{D} + \delta^r_C  g_{D\alpha}\right)
   \left( \epsilon^{CKLMN} \del_E \left( F^a_{KL}R_{MN}{}^{DE} \right) +\Gamma^C_{EB} \epsilon^{BKLMN} F^a_{KL}R_{MN}{}^{DE} \right)
\end{split}\\
\begin{split} \label{eq:GC_full_developed}
     & = 2 g_{\alpha r} \epsilon^{\kappa\lambda\mu\nu} \del_\epsilon \left(F^a_{\kappa\lambda} R_{\mu\nu}{}^{r\epsilon}\right)
     + 2g_{\alpha\sigma}\epsilon^{\sigma\kappa\mu\nu}\del_\epsilon \left( F^a_{\kappa r} R_{\mu\nu}{}^{r\epsilon}\right)
     + 2g_{\alpha\sigma}\epsilon^{\sigma\kappa\lambda\mu}\del_\epsilon \left( F^a_{\kappa \lambda} R_{\mu r}{}^{r\epsilon}\right)\\
     & + \epsilon^{\kappa\lambda\mu\nu}g_{\delta\alpha}\del_r\left( F^a_{\kappa\lambda} R_{\mu\nu}{}^{\delta r} \right)
     + \epsilon^{\kappa\lambda\mu\nu}g_{\delta\alpha}\del_\epsilon\left( F^a_{\kappa\lambda} R_{\mu\nu}{}^{\delta \epsilon} \right)\\
     & + \Gamma_{\alpha\epsilon r}\epsilon^{\kappa\lambda\mu\nu}F^a_{\kappa\lambda}R_{\mu\nu}{}^{r \epsilon}
     + 2 \Gamma_{\alpha\epsilon\beta}\epsilon^{\beta\kappa\mu\nu} F^a_{\kappa r} R_{\mu\nu}{}^{r \epsilon}
     + 2 \Gamma_{\alpha\epsilon\beta}\epsilon^{\beta\kappa\lambda\mu} F^a_{\kappa \lambda} R_{\mu r}{}^{r \epsilon}\\
     & + \Gamma^r_{\epsilon r} \epsilon^{\kappa \lambda \mu \nu} F^a_{\kappa\lambda} R_{\mu\nu\alpha}{}^\epsilon
     + 2 \Gamma^r_{r \beta} \epsilon^{\beta\kappa\mu\nu}F^a_{\kappa r} R_{\mu\nu\alpha}{}^r
     + 2\Gamma^r_{r \beta}\epsilon^{\beta\kappa\lambda\mu}F^a_{\kappa\lambda} R_{\mu r \alpha}{}^r \\
     & + 2 \Gamma^r_{\epsilon\beta} \epsilon^{\beta\kappa\mu\nu}F^a_{\kappa r}R_{\mu\nu\alpha}{}^\epsilon
     + 2 \Gamma^r_{\epsilon\beta} \epsilon^{\beta\kappa\lambda\mu}F^a_{\kappa \lambda}R_{\mu r \alpha}{}^\epsilon.
\end{split}
\end{align}

The necessary Christoffel symbols, Riemann components and field strength components are listed in appendix \ref{app:useful_summary}.

Since we are working up to second order in derivatives, the first term in \eqref{eq:GC_full_developed} drops immediately, using $R^{(0)}_{\mu\nu}{}^{r\epsilon}=0$ and $F^{a(0)}_{\kappa \lambda}=0$.

When contracted with $u^\alpha$, the other terms in the first two lines of \eqref{eq:GC_full_developed} also drop, since on the horizon we have $g^{(0)}_{\beta\alpha} u^\alpha = 0$.

We are left with the last three lines of \eqref{eq:GC_full_developed}, contracted with $u^\alpha$.
Let's have a look at each term separately:
\begin{align*}
\begin{split} 
     & u^\alpha\Gamma^{(0)}_{\alpha\epsilon r}\epsilon^{\kappa\lambda\mu\nu}F^{a(1)}_{\kappa\lambda}
     R^{(1)}_{\mu\nu}{}^{r \epsilon} =
     -\frac{1}{2}k'u_\epsilon \epsilon^{\kappa\lambda\mu\nu} F^{a(1)}_{\kappa\lambda} \left[-u^\epsilon \del_\mu(k'u_\nu)+P^{\epsilon\sigma} \left(\dots\right) \right] = \\
     &\qquad\qquad\qquad\qquad\quad = k'\del_\mu(k'u_\nu) \epsilon^{\kappa\lambda\mu\nu} \del_\kappa (\mu^a u_\lambda)=
     16\pi^2T\epsilon^{\kappa\lambda\mu\nu}\del_\kappa (\mu^a u_\lambda)\del_\mu(Tu_\nu)
\end{split}\\
\begin{split} 
     &2u^\alpha \Gamma^{(1)}_{\alpha\epsilon\beta}\epsilon^{\beta\kappa\mu\nu} F^{(0)a}_{\kappa r} R^{(1)}_{\mu\nu}{}^{r \epsilon} \sim \epsilon^{\beta\kappa\mu\nu} \Gamma^{(1)}_{\alpha\epsilon\beta} u_\kappa u^\alpha
     \left[-u^\epsilon \del_\mu(k'u_\nu)+u_\nu \left(\dots\right) \right] \sim
     \Gamma^{(1)}_{\alpha\epsilon\beta} u^\epsilon u^\alpha =0
\end{split}\\
\begin{split} 
     & 2u^\alpha \Gamma^{(1)}_{\alpha\epsilon\beta}\epsilon^{\beta\kappa\lambda\mu} F^{a(1)}_{\kappa \lambda} R^{(0)}_{\mu r}{}^{r \epsilon}\sim
     u^\alpha \Gamma^{(1)}_{\alpha\epsilon\beta}\epsilon^{\beta\kappa\lambda\mu}
     \left[ u_\mu u^\epsilon (\dots) +\delta^\epsilon_\mu (\dots) \right] \\&
     \qquad\qquad\qquad\qquad\qquad
     \sim
     \Gamma^{(1)}_{\alpha\epsilon\beta} u^\epsilon u^\alpha + \epsilon^{\beta\kappa\lambda\mu} \Gamma^{(1)}_{\alpha\mu\beta}=
     0
\end{split}\\
\begin{split} 
     & u^\alpha \epsilon^{\kappa \lambda \mu \nu} F^{a(1)}_{\kappa\lambda}
     \left( \Gamma^{r(0)}_{\epsilon r} R^{(1)}_{\mu\nu\alpha}{}^\epsilon+
     \Gamma^{r(1)}_{\epsilon r} R^{(0)}_{\mu\nu\alpha}{}^\epsilon \right)=
     u^\alpha \epsilon^{\kappa \lambda \mu \nu} F^{a(1)}_{\kappa\lambda}
     \Gamma^{r(0)}_{\epsilon r} R^{(1)}_{\mu\nu\alpha}{}^\epsilon+u^\alpha R^{(0)}_{\mu\nu\alpha}{}^\epsilon\left( \dots \right)
     \\&\qquad\qquad
     =-\frac{k'}{2}\left(u^\alpha u_\epsilon R^{(1)}_{\mu\nu\alpha}{}^\epsilon\right) \epsilon^{\kappa \lambda \mu \nu} F^{a(1)}_{\kappa\lambda} = k' \epsilon^{\kappa \lambda \mu \nu}  \del_\mu (k' u_\nu) \del_\kappa (\mu^a u_\lambda)
     \\ &\qquad\qquad
     =  16\pi^2T\epsilon^{\kappa\lambda\mu\nu}\del_\kappa (\mu^a u_\lambda)\del_\mu(Tu_\nu)
\end{split}\\
\begin{split} 
     & 2 u^\alpha \epsilon^{\beta\kappa\mu\nu} R^{(1)}_{\mu\nu\alpha}{}^r
     \left(\Gamma^{r(0)}_{r \beta} F^{a(1)}_{\kappa r} +\Gamma^{r(1)}_{r \beta} F^{a(0)}_{\kappa r} \right)
     \sim u^\alpha R^{(1)}_{\mu\nu\alpha}{}^r =0
\end{split}\\
\begin{split} 
     & 2 u^\alpha \Gamma^r_{r \beta}\epsilon^{\beta\kappa\lambda\mu}F^{a}_{\kappa\lambda} R_{\mu r \alpha}{}^r
     \sim u^\alpha R^{(0,1)}_{\mu r \alpha}{}^r = 0
\end{split}\\
\begin{split} 
     & 2  u^\alpha \Gamma^{r}_{\epsilon\beta} \epsilon^{\beta\kappa\mu\nu}
     F^a_{\kappa r}R_{\mu\nu\alpha}{}^\epsilon \sim
     u^\alpha R^{(0)}_{\mu\nu\alpha}{}^\epsilon \left(\dots\right)
     + u^\alpha \Gamma^{r(1)}_{\epsilon\beta} \epsilon^{\beta\kappa\mu\nu}
     F^{a(0)}_{\kappa r}R^{(1)}_{\mu\nu\alpha}{}^\epsilon  \\
     &\qquad\qquad
     \sim 0+u^\alpha \Gamma^{r(1)}_{\epsilon\beta} \epsilon^{\beta\kappa\mu\nu}
     u_\kappa R^{(1)}_{\mu\nu\alpha}{}^\delta \left( P_\delta^\epsilon-u_\delta u^\epsilon \right)
     \\
     &\qquad\qquad =
     \epsilon^{\beta\kappa\mu\nu} u_\kappa \left(R^{(1)}_{\mu\nu\alpha}{}^\delta u^\alpha  P_\delta^\epsilon \right) (\dots) +\Gamma^{r(1)}_{\epsilon\beta} u^\epsilon (\dots) = \epsilon^{\beta\kappa\mu\nu} u_\kappa u_\mu (\dots)   +0=0
\end{split}\\
\begin{split} 
     & 2 u^\alpha \Gamma^{r(1)}_{\epsilon\beta} \epsilon^{\beta\kappa\lambda\mu}F^{a(1)}_{\kappa \lambda}
     R^{(0)}_{\mu r \alpha}{}^\epsilon \sim
     u^\alpha \Gamma^{r(1)}_{\epsilon\beta} \epsilon^{\beta\kappa\lambda\mu}
     F^{a(1)}_{\kappa \lambda} \left(u^\epsilon (\dots) + \delta_\mu^\epsilon (\dots) \right)\\&\qquad\qquad
     = \Gamma^{r(1)}_{\epsilon\beta} u^\epsilon (\dots) +\epsilon^{\beta\kappa\lambda\mu} \Gamma^{r(1)}_{\mu\beta}=0 \ .
\end{split}
\end{align*}
Summing up, we eventually get:
\begin{align}
 \begin{split}
 &2\lambda_a u^\alpha \left(g_{\alpha C}\delta^{r}_{D} +
 \delta^r_C g_{D\alpha}\right)\epsilon^{CKLMN}\nabla_E\left(F^a_{KL}R_{MN}{}^{DE}\right)=
 64 \lambda_a\pi^2T\epsilon^{\kappa\lambda\mu\nu}\del_\kappa (\mu^a u_\lambda)\del_\mu(Tu_\nu) .
 \end{split}
\end{align}

This is the contribution to the focusing equation from the mixed gauge-gravitational Chern-Simons term.

\section{Chern-Simons contribution to the Gauss law} \label{app:Maxwell}

In this appendix, we derive the relation \eqref{eq:maxwell_to_check1}, i.e:
\begin{align}
 \epsilon^{\mu\nu\rho\sigma}R_{\mu\nu A}{}^B R_{\rho\sigma B}{}^A &= 32\pi^2\epsilon^{\mu\nu\rho\sigma}\del_\mu(Tu_\nu)\del_\rho(Tu_\sigma) \ . \label{eq:maxwell_to_check}
\end{align}
All metric and Riemann components needed for our calculation are listed in appendix \ref{app:useful_summary}.
We work up to second order in derivatives.

We can use the zeroth order Riemann components (eqs. \eqref{eq:sum_R_munuA_B_0}-\eqref{eq:sum_R_munu_AB_0})
to express $\epsilon^{\mu\nu\rho\sigma}R_{\mu\nu A}{}^B R_{\rho\sigma B}{}^A$ to second order without using the second-order Riemann tensor. We start by writing:
\begin{align}
 \begin{split}
   \epsilon^{\mu\nu\rho\sigma}R_{\mu\nu A}{}^B R_{\rho\sigma B}{}^A ={}& \epsilon^{\mu\nu\rho\sigma}R_{\mu\nu AB} R_{\rho\sigma}{}^{BA} \\
     ={}& \epsilon^{\mu\nu\rho\sigma}\left(R^{(0)}_{\mu\nu AB} R^{(0)}_{\rho\sigma}{}^{BA} + R^{(0)}_{\mu\nu AB} R^{(1)}_{\rho\sigma}{}^{BA}
      + R^{(1)}_{\mu\nu AB} R^{(0)}_{\rho\sigma}{}^{BA} \right. \\
     &\left.{} + R^{(0)}_{\mu\nu AB} R^{(2)}_{\rho\sigma}{}^{BA}
      + R^{(1)}_{\mu\nu AB} R^{(1)}_{\rho\sigma}{}^{BA} + R^{(2)}_{\mu\nu AB} R^{(0)}_{\rho\sigma}{}^{BA}\right) + O(\del^3) \ . \label{eq:epsilon_RR_raw}
 \end{split}
\end{align}
We see from \eqref{eq:sum_R_0_lower}-\eqref{eq:sum_R_munu_AB_0} that the contribution $R^{(0)}_{\mu\nu AB} R^{(0)}_{\rho\sigma}{}^{BA}$ vanishes. The $R^{(2)}_{\mu\nu AB} R^{(0)}_{\rho\sigma}{}^{BA}$ term also vanishes, as follows:
\begin{align}
 \begin{split}
   \epsilon^{\mu\nu\rho\sigma}R^{(2)}_{\mu\nu AB} R^{(0)}_{\rho\sigma}{}^{BA} &= \epsilon^{\mu\nu\rho\sigma}R^{(2)}_{\mu\nu\alpha\beta} R^{(0)}_{\rho\sigma}{}^{\beta\alpha}
    = \frac{f'k'}{f}\epsilon^{\mu\nu\rho\sigma}R^{(2)}_{\mu\nu\alpha\beta}u_\sigma u^\beta\delta_\rho^\alpha \\
    &= \frac{f'k'}{f}\epsilon^{\mu\nu\rho\sigma}R^{(2)}_{\mu\nu\rho\beta}u_\sigma u^\beta = 0 \ .
 \end{split}
\end{align}
In the last equality, we used the Bianchi identity. For the $R^{(0)}_{\mu\nu AB} R^{(2)}_{\rho\sigma}{}^{BA}$ contribution in \eqref{eq:epsilon_RR_raw}, we have:
\begin{align}
 \begin{split}
   & \epsilon^{\mu\nu\rho\sigma}R^{(0)}_{\mu\nu AB} R^{(2)}_{\rho\sigma}{}^{BA} = 2\epsilon^{\mu\nu\rho\sigma}R^{(0)}_{\mu\nu r\alpha} R^{(2)}_{\rho\sigma}{}^{\alpha r} \\
   &{}= 2\epsilon^{\mu\nu\rho\sigma}R^{(0)}_{\mu\nu r\alpha}\left(g^{(0)\alpha A} R^{(2)}_{\rho\sigma A}{}^{r} + g^{(1)\alpha A} R^{(1)}_{\rho\sigma A}{}^{r}
      + g^{(2)\alpha A} R^{(0)}_{\rho\sigma A}{}^{r}\right) \\
   &{}= 2\epsilon^{\mu\nu\rho\sigma}R^{(0)}_{\mu\nu r\alpha}\left(u^\alpha R^{(2)}_{\rho\sigma r}{}^{r} + \frac{1}{f}P^{\alpha\beta}R^{(2)}_{\rho\sigma\beta}{}^{r}
      + g^{(1)\alpha A} R^{(1)}_{\rho\sigma A}{}^{r} + 0 \right) \ .
 \end{split}
\end{align}
The first term vanishes, since $R^{(0)}_{\mu\nu r\alpha}u^\alpha = 0$. The second term also vanishes, as follows:
\begin{align}
 \epsilon^{\mu\nu\rho\sigma}R^{(0)}_{\mu\nu r\alpha}P^{\alpha\beta}R^{(2)}_{\rho\sigma\beta}{}^{r}
  = -\frac{1}{2}f'k'\epsilon^{\mu\nu\rho\sigma}u_\mu\delta_\nu^\beta R^{(2)}_{\rho\sigma\beta}{}^{r}
  = -\frac{1}{2}f'k'\epsilon^{\mu\nu\rho\sigma}u_\mu R^{(2)}_{\rho\sigma\nu}{}^{r} = 0 \ ,
\end{align}
where we again used the Bianchi identity. Putting everything together, we are left with the expression:
\begin{align}
 \begin{split}
   \epsilon^{\mu\nu\rho\sigma}R_{\mu\nu A}{}^B R_{\rho\sigma B}{}^A ={}& \epsilon^{\mu\nu\rho\sigma}\left(R^{(0)}_{\mu\nu}{}^{AB} R^{(1)}_{\rho\sigma BA}
     + R^{(0)}_{\mu\nu AB} R^{(1)}_{\rho\sigma}{}^{BA} \right. \\
     &\left.{} + 2R^{(0)}_{\mu\nu r\alpha} g^{(1)\alpha A} R^{(1)}_{\rho\sigma A}{}^{r}
     + R^{(1)}_{\mu\nu AB} R^{(1)}_{\rho\sigma}{}^{BA}\right) + O(\del^3) \ . \label{eq:epsilon_RR_no_R2}
 \end{split}
\end{align}
We can now bring the first-order Riemann tensor to the standard index placement $R_{\mu\nu A}{}^B$. The indices of $R^{(0)}_{\mu\nu A}{}^B$ can be raised and lowered with the zeroth-order metric. When raising or lowering $R^{(1)}_{\mu\nu A}{}^B$, we get expressions of the form $g^{(0)}R^{(1)} + g^{(1)}R^{(0)}$ (with indices suppressed). Whenever an $R^{(0)}$ factor is multiplied by another $R^{(0)}$, it can be dropped, since $\epsilon^{\mu\nu\rho\sigma}R^{(0)}_{\mu\nu A}{}^B R^{(0)}_{\rho\sigma C}{}^D = 0$ due to the antisymmetric product of two $u_\mu$'s. In this way, the first two terms in \eqref{eq:epsilon_RR_no_R2} can be written as $2\epsilon^{\mu\nu\rho\sigma}R^{(0)}_{\mu\nu A}{}^B R^{(1)}_{\rho\sigma B}{}^A$. For the last term, we have:
\begin{align}
 \epsilon^{\mu\nu\rho\sigma}R^{(1)}_{\mu\nu AB} R^{(1)}_{\rho\sigma}{}^{BA} = \epsilon^{\mu\nu\rho\sigma}
   \left(R^{(1)}_{\mu\nu A}{}^C g^{(0)}_{BC} + R^{(0)}_{\mu\nu A}{}^C g^{(1)}_{BC}\right)\left(R^{(1)}_{\rho\sigma D}{}^A g^{(0)BD} + R^{(0)}_{\rho\sigma D}{}^A g^{(1)BD}\right) .
\end{align}
Using $g^{(0)}_{AB}g^{(0)BC} = \delta_A^C$ and $g^{(0)}_{AB}g^{(1)BC} + g^{(1)}_{AB}g^{(0)BC} = 0$, this can be written as:
\begin{align}
 \epsilon^{\mu\nu\rho\sigma}R^{(1)}_{\mu\nu AB} R^{(1)}_{\rho\sigma}{}^{BA} = \epsilon^{\mu\nu\rho\sigma}\left(R^{(1)}_{\mu\nu A}{}^B R^{(1)}_{\rho\sigma B}{}^{A}
   - R^{(0)}_{\mu\nu AB}\, g^{(1)BC} R^{(1)}_{\rho\sigma C}{}^A - R^{(0)}_{\mu\nu}{}^{BA} g^{(1)}_{BC} R^{(1)}_{\rho\sigma A}{}^C \right) .
\end{align}
Putting things together again, we end up with:
\begin{align}
 \begin{split}
   \epsilon^{\mu\nu\rho\sigma}R_{\mu\nu A}{}^B R_{\rho\sigma B}{}^A ={}&
   \epsilon^{\mu\nu\rho\sigma}\left(   2R^{(0)}_{\mu\nu A}{}^B R^{(1)}_{\rho\sigma B}{}^A
       + 2R^{(0)}_{\mu\nu r\alpha} g^{(1)\alpha A} R^{(1)}_{\rho\sigma A}{}^{r}  \right. \\
       &\left.{} - R^{(0)}_{\mu\nu AB}\, g^{(1)BC} R^{(1)}_{\rho\sigma C}{}^A
       - R^{(0)}_{\mu\nu}{}^{BA} g^{(1)}_{BC} R^{(1)}_{\rho\sigma A}{}^C +
       R^{(1)}_{\mu\nu A}{}^B R^{(1)}_{\rho\sigma B}{}^A \right) \\
     ={}& \epsilon^{\mu\nu\rho\sigma}\left(2R^{(0)}_{\mu\nu A}{}^\beta R^{(1)}_{\rho\sigma\beta}{}^A + R^{(0)}_{\mu\nu r\alpha} g^{(1)\alpha A} R^{(1)}_{\rho\sigma A}{}^{r}
        \right. \\
       &\left.{} - R^{(0)}_{\mu\nu\alpha r}\, g^{(1)rC} R^{(1)}_{\rho\sigma C}{}^\alpha
       - R^{(0)}_{\mu\nu}{}^{\beta\alpha} g^{(1)}_{\beta C} R^{(1)}_{\rho\sigma\alpha}{}^C + R^{(1)}_{\mu\nu A}{}^B R^{(1)}_{\rho\sigma B}{}^A \right) \\
     ={}& \epsilon^{\mu\nu\rho\sigma}\left(2R^{(0)}_{\mu\nu A}{}^\beta R^{(1)}_{\rho\sigma\beta}{}^A
       + R^{(0)}_{\mu\nu r\alpha}\left(g^{(1)\alpha A} R^{(1)}_{\rho\sigma A}{}^{r} + g^{(1)rA} R^{(1)}_{\rho\sigma A}{}^\alpha \right) \right. \\
       &\left.{} - R^{(0)}_{\mu\nu}{}^{\beta\alpha} g^{(1)}_{\beta A} R^{(1)}_{\rho\sigma\alpha}{}^A + R^{(1)}_{\mu\nu A}{}^B R^{(1)}_{\rho\sigma B}{}^A \right) \ . \label{eq:epsilon_RR}
 \end{split}
\end{align}
The Riemann components $R^{(1)}_{\mu\nu A}{}^B$ can be found in eqs. \eqref{eq:sum_Riemann_r_dep_1}-\eqref{eq:sum_Riemann_1_decomp} of appendix \ref{app:useful_summary}.

We are now ready to evaluate the expression \eqref{eq:epsilon_RR}. For the $R^{(0)}_{\mu\nu A}{}^\beta R^{(1)}_{\rho\sigma\beta}{}^A$ term, the relevant components of $R^{(1)}_{\rho\sigma\beta}{}^A$ are $R^{(1)}_{\rho\sigma\beta}{}^r$ and $R^{(1)}_{\rho\sigma\beta}{}^\gamma u^\beta P^\alpha_\gamma$. Their product with $\epsilon^{\mu\nu\rho\sigma}R^{(0)}_{\mu\nu A}{}^\beta$ vanishes, due to the antisymmetrized product of two $u_\mu$'s. For the second term in \eqref{eq:epsilon_RR}, we have:
\begin{align}\label{app:second_term}
 \epsilon^{\mu\nu\rho\sigma}R^{(0)}_{\mu\nu r\alpha}\left(g^{(1)\alpha A} R^{(1)}_{\rho\sigma A}{}^{r} + g^{(1)rA} R^{(1)}_{\rho\sigma A}{}^\alpha \right)
  = \epsilon^{\mu\nu\rho\sigma}R^{(0)}_{\mu\nu r\alpha}\left(-\frac{1}{f^2}
  h^{\alpha\kappa} g^{(1)}_{\kappa\lambda} h^{\lambda\beta} R^{(1)}_{\rho\sigma\beta}{}^{r} + 0 \right) \ ,
\end{align}
where $h_{\mu\nu}$ is the fluid (flat) metric.
This term again vanishes due to an antisymmetrized product of two $u_\mu$'s.
For the third term in \eqref{eq:epsilon_RR}, we have:
\begin{align}
 \epsilon^{\mu\nu\rho\sigma}R^{(0)}_{\mu\nu}{}^{\beta\alpha} g^{(1)}_{\beta A} R^{(1)}_{\rho\sigma\alpha}{}^A
   = \epsilon^{\mu\nu\rho\sigma}R^{(0)}_{\mu\nu}{}^{\beta\alpha} g^{(1)}_{\beta\gamma} R^{(1)}_{\rho\sigma\alpha}{}^\gamma
   = \frac{f'k'}{2f}\epsilon^{\mu\nu\rho\sigma}u_\mu u^\alpha g^{(1)}_{\nu\gamma} R^{(1)}_{\rho\sigma\alpha}{}^\gamma \ .
\end{align}
Again the relevant components are $R^{(1)}_{\rho\sigma\alpha}{}^\gamma u^\alpha P^\beta_\gamma$, whose product with $\epsilon^{\mu\nu\rho\sigma}u_\mu$ vanishes. We are left with the product $\epsilon^{\mu\nu\rho\sigma}R^{(1)}_{\mu\nu A}{}^B R^{(1)}_{\mu\nu B}{}^A$. This decomposes as:
\begin{align}
 \begin{split}
   \epsilon^{\mu\nu\rho\sigma}R^{(1)}_{\mu\nu A}{}^B R^{(1)}_{\mu\nu B}{}^A ={}& \epsilon^{\mu\nu\rho\sigma}\left(R^{(1)}_{\mu\nu r}{}^r R^{(1)}_{\rho\sigma r}{}^r
     + 2R^{(1)}_{\mu\nu r}{}^\alpha R^{(1)}_{\rho\sigma\alpha}{}^r + R^{(1)}_{\mu\nu\alpha}{}^\beta u^\alpha u_\beta R^{(1)}_{\rho\sigma\gamma}{}^\delta u^\gamma u_\delta
     \right.\\
     &\left.{} - 2R^{(1)}_{\mu\nu\alpha}{}^\beta u^\alpha P^\gamma_\beta R^{(1)}_{\rho\sigma\gamma}{}^\delta u_\delta
     + R^{(1)}_{\mu\nu\alpha}{}^\beta P_\delta^\alpha P^\gamma_\beta R^{(1)}_{\rho\sigma\gamma}{}^\delta \right) \ .
 \end{split}
\end{align}
The $\epsilon^{\mu\nu\rho\sigma}R^{(1)}_{\mu\nu r}{}^r R^{(1)}_{\rho\sigma r}{}^r$ and $\epsilon^{\mu\nu\rho\sigma}R^{(1)}_{\mu\nu\alpha}{}^\beta u^\alpha u_\beta R^{(1)}_{\rho\sigma\gamma}{}^\delta u^\gamma u_\delta$ terms contribute $\epsilon^{\mu\nu\rho\sigma}\del_\mu(k' u_\nu)\del_\rho(k' u_\sigma)$ each. The other terms don't contribute, as we now show:
\begin{align}
 \begin{split}
   \epsilon^{\mu\nu\rho\sigma}R^{(1)}_{\mu\nu r}{}^\alpha R^{(1)}_{\rho\sigma\alpha}{}^r
     ={}& \epsilon^{\mu\nu\rho\sigma}\left(u_\mu\left(\dots\right) + \delta_\mu^\alpha\left(\dots\right)\right) R^{(1)}_{\rho\sigma\alpha}{}^r \\
     ={}& \epsilon^{\mu\nu\rho\sigma}\left(u_\mu u_\rho\left(\dots\right) + R^{(1)}_{\rho\sigma\mu}{}^r\left(\dots\right) \right) = 0 + 0 = 0
 \end{split} \\
 \begin{split}
   \epsilon^{\mu\nu\rho\sigma}R^{(1)}_{\mu\nu\alpha}{}^\beta u^\alpha P^\gamma_\beta R^{(1)}_{\rho\sigma\gamma}{}^\delta u_\delta
     ={}& \epsilon^{\mu\nu\rho\sigma}R^{(1)}_{\mu\nu\alpha}{}^\beta u^\alpha P^\gamma_\beta \left(u_\rho\left(\dots\right) + P_{\rho\gamma}\left(\dots\right) \right) \\
     ={}& \epsilon^{\mu\nu\rho\sigma}\left(u_\mu u_\rho\left(\dots\right) + R^{(1)}_{\mu\nu\alpha}{}^\beta u^\alpha P_{\beta\rho}\left(\dots\right) \right) \\
     ={}& 0 + k'\epsilon^{\mu\nu\rho\sigma}u_\mu\left(\frac{1}{2f}P_{\nu\rho} u^\lambda\del_\lambda f + P^\lambda_\rho\del_{(\lambda}u_{\nu)} \right)\left(\dots\right) \\
     ={}& 0 + k'\epsilon^{\mu\nu\rho\sigma}u_\mu\left(\del_{(\nu}u_{\rho)} + \frac{1}{2}u_\rho u^\lambda\del_\lambda u_\nu \right)\left(\dots\right)
     = 0 + 0 = 0
 \end{split} \\
 \begin{split}
   \epsilon^{\mu\nu\rho\sigma}R^{(1)}_{\mu\nu\alpha}{}^\beta P_\delta^\alpha P^\gamma_\beta R^{(1)}_{\rho\sigma\gamma}{}^\delta
     ={}& \epsilon^{\mu\nu\rho\sigma}\left(P_{\mu\alpha}\left(\dots\right) + P_\mu^\beta\left(\dots\right)\right)
         \left(P_{\rho\beta}\left(\dots\right) + P_\rho^\alpha\left(\dots\right)\right) \\
     ={}& \epsilon^{\mu\nu\rho\sigma}\left(P_{\mu\rho}\left(\dots\right) + P_{\mu\alpha}P_{\rho\beta}\left(\dots\right) + P_\mu^\beta P_\rho^\alpha\left(\dots\right)\right) \\
     ={}& 0 + \epsilon^{\mu\nu\rho\sigma}\left(f'^2 P^\kappa_\mu P^\lambda_\rho P_\nu^\gamma P_\sigma^\delta\del_{(\lambda}u_{\gamma)}\del_{(\kappa}u_{\delta)} \right. \\
     &\left.{} + P_\mu^\beta P_\rho^\alpha\left(P_{\nu\alpha}u^\lambda\del_\lambda f
       + f P_\nu^\kappa P_\alpha^\lambda\del_{(\kappa}u_{\lambda)} \right) \left(\dots\right) \right) \\
     ={}& 0 + \epsilon^{\mu\nu\rho\sigma}P_\mu^\beta \left(P_{\nu\rho}u^\lambda\del_\lambda f + f P_\nu^\kappa P_\rho^\lambda\del_{(\kappa}u_{\lambda)} \right) \left(\dots\right)
     = 0 \ .
 \end{split}
\end{align}
We finally get:
\begin{align}
 \begin{split}
   \epsilon^{\mu\nu\rho\sigma}R^{(1)}_{\mu\nu A}{}^B R^{(1)}_{\mu\nu B}{}^A &= \epsilon^{\mu\nu\rho\sigma}\left(R^{(1)}_{\mu\nu r}{}^r R^{(1)}_{\mu\nu r}{}^r
       + R^{(1)}_{\mu\nu\alpha}{}^\beta u^\alpha u_\beta R^{(1)}_{\rho\sigma\gamma}{}^\delta u^\gamma u_\delta \right) \\
     &= 2\epsilon^{\mu\nu\rho\sigma}\del_\mu(k' u_\nu)\del_\rho(k' u_\sigma) = 32\pi^2\epsilon^{\mu\nu\rho\sigma}\del_\mu(Tu_\nu)\del_\rho(Tu_\sigma)
 \end{split} \\
 \begin{split}
   \epsilon^{\mu\nu\rho\sigma}R_{\mu\nu A}{}^B R_{\mu\nu B}{}^A &= \epsilon^{\mu\nu\rho\sigma}R^{(1)}_{\mu\nu A}{}^B R^{(1)}_{\mu\nu B}{}^A + O(\del^3) \\
     &= 32\pi^2\epsilon^{\mu\nu\rho\sigma}\del_\mu(Tu_\nu)\del_\rho(Tu_\sigma) + O(\del^3) \ .
 \end{split}
\end{align}
Eq. \eqref{eq:maxwell_to_check} is therefore demonstrated.

\section{Chern-Simons contribution to the Gauss-Codazzi equation} \label{app:GC}

In this appendix, we derive eq. \eqref{eq:GC_to_check1}, i.e.:
\begin{align}
 \begin{split}
   \left(g_{\mu C}g^{rD} + \delta^r_C\delta^D_\mu\right) & \epsilon^{CKLMN}\nabla_E \left(F^a_{KL}R_{MND}{}^E\right)
     \\& = -32\pi^2 Tu_\mu\epsilon^{\nu\rho\sigma\lambda}\del_\nu(\mu^a u_\rho)\del_\sigma(Tu_\lambda)
     - 8 s u^\nu \del_{[\mu}\left(\frac{Q^a}{s} \omega_{\nu]} \right) \ ,
 \end{split} \label{eq:GC_to_check_app}
\end{align}
where $Q^a \equiv \left( a^a(fk''-f'k')+a'^a f  k'\right) = f^2(k'a^a/f)'$.

In appendix \ref{app:extra_focusing}, eq. \eqref{eq:GC_full_developed}, we have already written the full expression for the LHS of \eqref{eq:GC_to_check_app}.

Let us now use the Christoffel symbols, Riemann components and field strength components from appendix \ref{app:useful_summary} to evaluate each term in \eqref{eq:GC_full_developed} separately (this time without projecting their vector index).

Up to second order in derivatives, we have:
\begin{align*}
\begin{split} 
    & 2 g_{\alpha r} \epsilon^{\kappa\lambda\mu\nu} \del_\epsilon \left(F^a_{\kappa\lambda} R_{\mu\nu}{}^{r\epsilon}\right) = O(\del^3) \equiv 0
\end{split} \\
\begin{split} 
    &
    2g_{\alpha\sigma}\epsilon^{\sigma\kappa\mu\nu}\del_\epsilon \left( F^a_{\kappa r} R_{\mu\nu}{}^{r\epsilon}\right)=
    4f\left[ \omega_\alpha \del_\epsilon (u^\epsilon a'^a k')
    +a'^a k' \left(u^\epsilon \del_\epsilon \omega_\alpha - \frac{1}{2}u_\alpha \del_\epsilon \omega^\epsilon \right)\right]
\end{split}\\
\begin{split} 
     &
     2g_{\alpha\sigma}\epsilon^{\sigma\kappa\lambda\mu}\del_\epsilon \left( F^a_{\kappa \lambda} R_{\mu r}{}^{r\epsilon}\right)=
     8f \omega_\alpha u^\epsilon\del_\epsilon  \left( a^a \left(\frac{k''}{2}-\frac{f'k'}{4f}\right) \right)
     \\& \qquad
     -4f P_{\alpha \sigma} \epsilon^{\sigma\kappa\lambda\mu} \del_\kappa \left( a^a u_\lambda \right) \del_\mu \left(\frac{f'k'}{4f} \right)
     \\& \qquad
     +  \left(2 k'' f - f'k'\right) a^a \left[2\omega_\alpha \del_\epsilon u^\epsilon +
        2u^\epsilon \del_\epsilon \omega_\alpha -u_\alpha \del_\epsilon \omega^\epsilon \right]
\end{split}\\
\begin{split} 
     & \epsilon^{\kappa\lambda\mu\nu}g_{\delta\alpha}\del_r\left( F^a_{\kappa\lambda} R_{\mu\nu}{}^{\delta r} \right)=
     4f\omega^\epsilon \del_\epsilon u_\alpha \left[(a^a k')'-a^a\frac{f'k'}{f} \right]\\
        & \qquad
        +2f \omega_\alpha\left[\left(\frac{a^ak'}{f}u^\alpha \del_\alpha f\right)' -\frac{a^a k' f'}{f^2}u^\alpha \del_\alpha f \right]\\
        & \qquad
     +\epsilon^{\kappa\lambda\mu\nu} \del_\kappa (a^a u_\lambda) P_{\mu\alpha}
        \left(-\frac{2k'f'}{f}\del_\nu f +f'\del_\nu k' +2k' \del_\nu f' \right)
\end{split}\\
\begin{split} 
     & \epsilon^{\kappa\lambda\mu\nu}g_{\alpha\delta}\del_\epsilon\left( F^a_{\kappa\lambda} R_{\mu\nu}{}^{\delta \epsilon} \right)=
     -2f\omega_\alpha \del_\epsilon \left(u^\epsilon\frac{a^af'k'}{f}\right)
     +2a^a f' k' \omega^\epsilon \del_\epsilon u_\alpha
     -a^af' k' (2u^\epsilon \del_\epsilon \omega_\alpha - u_\alpha \del_\epsilon \omega^\epsilon)
\end{split}\\
\begin{split} 
     & \Gamma_{\alpha\epsilon r}\epsilon^{\kappa\lambda\mu\nu}F^a_{\kappa\lambda}R_{\mu\nu}{}^{r \epsilon}=
     k' u_\alpha \epsilon^{\kappa\lambda\mu\nu} \del_\kappa(a^au_\lambda) \del_\mu(k' u_\nu)
     - 2a^a f' k' \left( \omega^\epsilon \del_\epsilon u_\alpha +\frac{1}{2f}\omega_\alpha u^\epsilon\del_\epsilon f\right)
\end{split}\\
\begin{split} 
     & 2 \Gamma_{\alpha\epsilon\beta}\epsilon^{\beta\kappa\mu\nu} F^a_{\kappa r} R_{\mu\nu}{}^{r \epsilon}=
     2 a'^a k' \omega_\alpha u_\epsilon \del^\epsilon f
\end{split}\\
\begin{split} 
     & 2 \Gamma_{\alpha\epsilon\beta}\epsilon^{\beta\kappa\lambda\mu} F^a_{\kappa \lambda} R_{\mu r}{}^{r \epsilon}=
     2 a^a \left(k''-\frac{f'k'}{2f}\right) \omega_\alpha u_\epsilon \del^\epsilon f
\end{split}\\
\begin{split} 
     &\Gamma^r_{\epsilon r} \epsilon^{\kappa \lambda \mu \nu} F^a_{\kappa\lambda} R_{\mu\nu\alpha}{}^\epsilon  =
      \frac{1}{2}f'k'^2 \epsilon^{\kappa\lambda\mu\nu} P_{\nu\alpha} u_\mu \del_\kappa A_\lambda^{a(0,1)}
      +k'u_\alpha \epsilon^{\kappa\lambda\mu\nu} \del_\mu \left( k'u_\nu \right) \del_\kappa \left(a^a u_\lambda \right)
      \\& \qquad
        +2a^ak'f' \omega^\epsilon \del_\epsilon u_\alpha
        +k' \epsilon^{\kappa\lambda\mu\nu} \del_\kappa \left(a^a u_\lambda \right)
        \biggr\{
        -f P_{\mu\alpha} P_\nu^\sigma \del_\sigma \left( \frac{f'}{f}\right)
        \\& \qquad
        + P_{\mu\alpha}u_\nu \sqrt{f} u^\sigma \del_\sigma \left(\frac{f'}{\sqrt{f}}\right)
        +\frac{f'}{2} P_{\mu\alpha} \left[a_\nu+ g'^{(1)}_{\nu\sigma}u^\sigma \right]
        + u_\mu \left[ \frac{k'}{2} P^\sigma_\alpha g'^{(1)}_{\sigma\nu}
        + \frac{f'}{2} P_{\nu\alpha} u^\sigma g'^{(1)}_{\sigma\epsilon} u^\epsilon \right]
        \biggr\}\
\end{split}\\
\begin{split} 
     & 2 \Gamma^r_{r \beta} \epsilon^{\beta\kappa\mu\nu}F^a_{\kappa r} R_{\mu\nu\alpha}{}^r=0
\end{split}\\
\begin{split} 
     & 2\Gamma^r_{r \beta}\epsilon^{\beta\kappa\lambda\mu}F^a_{\kappa\lambda} R_{\mu r \alpha}{}^r=
     -\frac{1}{2}f'k'^2 \epsilon^{\beta \kappa \lambda \mu} P_{\mu\alpha} u_\beta \del_\kappa A_\lambda^{a(0,1)}
     -  2k' f' a^a \omega^\epsilon \del_\epsilon u_\alpha
     + k' \epsilon^{\beta \kappa \lambda \mu} \del_\kappa \left( a^a u_\lambda \right) \times
     \\& \quad
     \left( u_\beta \left[
        \sqrt{f} P_{\mu\alpha} u^\sigma \del_\sigma \left(\frac{f'}{\sqrt{f}}\right)
        -\frac{k'}{2}P_\alpha^\sigma g'^{(1)}_{\sigma\mu} \right]
        +\frac{f'}{2} P_{\mu\alpha}
        \left[a_\beta+ g'^{(1)}_{\beta\delta}u^\delta
        -u_\beta u^{\delta} g^{'(1)}_{\delta\sigma} u^{\sigma} \right]
        \right)
\end{split}\\
\begin{split} 
     & 2 \Gamma^r_{\epsilon\beta} \epsilon^{\beta\kappa\mu\nu}F^a_{\kappa r}R_{\mu\nu\alpha}{}^\epsilon=0
\end{split}\\
\begin{split} 
     & 2 \Gamma^r_{\epsilon\beta} \epsilon^{\beta\kappa\lambda\mu}F^a_{\kappa \lambda}R_{\mu r \alpha}{}^\epsilon=0 \ ,
\end{split}
\end{align*}
where we have used the following definitions:
\begin{align}
  & \mbox{acceleration vector}: &  a_\mu & \equiv u^\nu \del_\nu u_\mu \ , \\
  & \mbox{vorticity vector}:  & \omega^\mu &\equiv\frac{1}{2}\epsilon^{\mu\nu\rho\sigma} u_\nu \del_\rho u_\sigma \ ,
\end{align}
as well as some geometrical identities (explained in section V of \cite{Eling:2010hu}):
\begin{align}
\begin{split}
&P_{\alpha\sigma} \epsilon^{\sigma\kappa\mu\nu}a_\kappa\del_\mu u_\nu = 0\\
& u^\nu\del_\mu \omega_\nu  =-\omega^\nu \del_\mu u_\nu \\
& a^\nu \omega_\nu = \frac{1}{2} \del_\nu \omega^\nu\\
& u^\nu\del_\nu \omega_\mu = u_\mu a_\nu \omega^\nu +\frac{1}{2}\epsilon_{\mu\alpha\beta\gamma} u^\alpha u^\nu \del_\nu \del^\beta u^\gamma\\
& \omega^\mu \del_\mu u_\alpha = \omega^\mu P_{\alpha\nu} \del^\nu u_\mu \ .
\end{split}
\end{align}
Summing all the terms, we get:
\begin{align}
     & 2 k'u_\alpha \epsilon^{\kappa\lambda\mu\nu} \del_\kappa \left(a^a u_\lambda \right)\notag
      \del_\mu \left( k'u_\nu \right)\\&\qquad
     +4 Q^a u^\epsilon \del_\epsilon \omega_\alpha  - 2Q^a u_\alpha \del_\epsilon \omega^\epsilon
     +4Q^a \omega^\epsilon \del_\epsilon u_\alpha + 4Q^a\omega_\alpha \del_\epsilon u^\epsilon
     +4\omega_\alpha u_\epsilon\del^\epsilon Q^a \notag
     \\& =
      -32\pi^2 Tu_\alpha \epsilon^{\kappa\lambda\mu\nu} \del_\kappa \left(\mu^a u_\lambda \right)
      \del_\mu \left( Tu_\nu \right)
      - 8 s u^\nu \del_{[\mu}\left(\frac{Q^a}{s} \omega_{\nu]} \right) \ ,
\end{align}
where we defined $Q^a(x^\mu) \equiv \left( a^a(fk''-f'k')+a'^a f  k'\right) = f^2(k'a^a/f)'$, and used
the definitions \eqref{eq:hydro_relate} of the thermal parameters.
We also used the the first-order focusing equation $\del_\mu (s u^\mu)=0$ (entropy conservation) for the last equality.
Eq. \eqref{eq:GC_to_check_app} is therefore demonstrated.

\section{Chern-Simons corrections to the hydrodynamic constitutive relations} \label{app:new_constit}

In this appendix, we derive the contributions \eqref{eq:new_constit} to the charge currents and the stress tensor from the $\lambda_a$ terms in the Einstein-Maxwell equations
\eqref{eq:maxwell_to_check2}-\eqref{eq:GC_to_check2}.

Using the equivalence \eqref{eq:equivalence} to relate the new contributions to conservation laws, one gets:
\begin{align}
 \begin{split}
   -\frac{1}{8\pi}\delta E_\mu^r + \frac{1}{4\pi}\mu^a u_\mu \delta M_a^r \quad &\Rightarrow \quad
    \del_\nu \left( \delta T^\nu_\mu \right) = 0 \\
   \frac{1}{4\pi}\delta M_a^r \quad &\Rightarrow \quad \del_\mu \left( \delta J_a^\mu \right) = 0 \ .
 \end{split}
\end{align}
Then using the explicit form of the new contributions \eqref{eq:maxwell_to_check2}-\eqref{eq:GC_to_check2} (remember those were on the RHS of the field equations), one gets:
\begin{align}
 \begin{split}
   - 8\pi \lambda_a \biggr\{2 Tu_\mu \del_\nu(\mu^a T \omega^\nu)
   - 4 s u^\nu \del_{[\mu}\left(\frac{\mu^a T^2}{s} \omega_{\nu]} \right)
+  \mu^a u_\mu \del_\nu( T^2 \omega^\nu) \biggr\} \quad &\Rightarrow \quad \del_\nu T^\nu_\mu = 0 \\
   -\del_\mu(8\pi \lambda_a T^2 \omega^\mu) \quad &\Rightarrow \quad \del_\mu J_a^\mu = 0 \ ,
 \end{split}
\end{align}
where we have used
\begin{align}
\epsilon^{\mu\nu\rho\sigma}\del_\mu(Tu_\nu)\del_\rho(Tu_\sigma)& = 2\del_\mu(T^2 \omega^\mu) \\
\epsilon^{\mu\nu\rho\sigma}\del_\mu(\mu^a u_\nu)\del_\rho(Tu_\sigma)& = 2\del_\mu(\mu^a T \omega^\mu) \ .
\end{align}
For the stress-energy tensor, some more work is required.
Let us develop it a bit:
\begin{align}
&- 8\pi \lambda_a \biggr\{2 Tu_\mu \del_\nu(\mu^a T \omega^\nu)
    +\mu^a u_\mu \del_\nu( T^2 \omega^\nu)
   - 4 s u^\nu \del_{[\mu}\left(\frac{\mu^a T^2}{s} \omega_{\nu]} \right) \biggr\}=
\\
&- 8\pi \lambda_a \biggr\{3\del_\nu(\mu^aT^2u_\mu \omega^\nu) - 2\del_\nu(Tu_\mu) \mu^a T \omega^\nu - \del_\nu(\mu^a u_\mu) T^2 \omega^\nu\\
& \qquad\qquad - 4u^\nu \del_{[\mu} (\mu^a T^2 \omega_{\nu]})
+ 2s u^\nu \mu^a T^2  \omega_{\mu}\del_{\nu}\left(\frac{1}{s}\right)\biggr\} \ .
\end{align}
Using the first-order focusing equation $\del_\mu(su^\mu)$ (see text above eq. \eqref{eq:focusing_2}) and remembering we are only interested in second order in derivatives, this becomes:
\begin{align*}
 = &- 8\pi \lambda_a
 \biggr\{3\del_\nu(\mu^aT^2u_\mu \omega^\nu) - 3 \omega^\nu \del_\nu u_\mu \mu^a T^2
 - \del_\nu(\mu^a T^2) u_\mu \omega^\nu\\
 & +2 \del_{\nu} (\mu^a T^2 u^\nu \omega_{\mu})
 + 4\mu^a T^2 \omega_{[\nu}  \del_{\mu]} u^\nu
 + 2 \mu^a T^2  \omega_{\mu} \del_\nu u^\nu\biggr\}
 \\
 =&- 8\pi \lambda_a
 \biggr\{4\del^\nu(\mu^aT^2u_{(\mu} \omega_{\nu)})+\del_\nu(\mu^aT^2u_\mu \omega^\nu)
 \\
 &  - 2 \omega^\nu \del_\nu u_\mu \mu^a T^2
 -  \omega^\nu \del_\nu ( \mu^a T^2 u_\mu)
 + 2\mu^a T^2 \omega_{\nu}  \del_{\mu} u^\nu
 \biggr\}
 \\
 =&- 8\pi \lambda_a
 \biggr\{4\del^\nu(\mu^aT^2u_{(\mu} \omega_{\nu)})+\del_\nu(\mu^aT^2u_\mu \omega^\nu)
 - 4 \mu^a T^2 \omega^\nu \del_{[\nu} u_{\mu]}
 -  \omega^\nu \del_\nu ( \mu^a T^2 u_\mu)\biggr\} \ .
\end{align*}
Using the hydrodynamic identity:
\begin{equation}
\omega^\nu \del_{[\nu} u_{\mu]} = \frac{1}{2} u_\mu \omega^\nu a_\nu = \frac{1}{4} u_\mu \del_\nu \omega^\nu \ ,
\end{equation}
where $a_\mu=u^\nu\del_\nu u_\mu$ is the acceleration vector,
only the first term survives.
We finally get:
\begin{align*}
- 32\pi \lambda_a \del^\nu(\mu^aT^2u_{(\mu} \omega_{\nu)}) \ ,
\end{align*}
ending up with:
\begin{align}
 \delta T^{\mu\nu} &= -32\pi\lambda_a\mu^a T^2 u^{(\mu}\omega^{\nu)} \\
 \delta J_a^\mu &= -8\pi\lambda_a T^2\omega^\mu
\end{align}
We have thus demonstrated eqs. \eqref{eq:new_constit}.

\section{List of Christoffel symbols, Riemann components and field strength components} \label{app:useful_summary}

We now summarize the Christoffel symbols, Riemann components and field strength components which we need for the calculations in the previous appendices.

As mentioned in sections \ref{sec:pre:hydro} and \ref{sec:chern:current}, we use the following hydrodynamic ansatz for the bulk fields (metric and gauge potential) in the neighborhood of the horizon:
\begin{align}
\begin{split}
  &g_{rr} = 0;\quad g_{r\mu} = -u_\mu;\quad g^{(0)}_{\mu\nu} = f(r,x^\mu) P_{\mu\nu} + k(r,x^\mu) u_\mu u_\nu;
  \\
  & A^{a}_{r} = 0; \quad A^{a(0)}_{\mu} = a^a(r,x^\mu) u_\mu \ ,
\end{split}
\end{align}
with first order corrections to the metric $g^{(1)}_{\mu\nu}$ satisfying $g^{(1)}_{\mu\nu} u^\nu=0$ on the horizon, and first order corrections $A^{a(1)}_{\mu}$ to the gauge potential.

On the horizon $g_{\mu\nu}$ becomes degenerate, i.e. $k(0) = 0$. The other functions and their radial derivatives are related (at zeroth order) to thermal parameters as follows (cf. \ref{sec:pre:hydro}, \ref{sec:chern:current}):
\begin{align}\label{eq:hydro_relate}
 f(0) = (4s)^{2/3};\quad k'(0) = -4\pi T;\quad a^a(0) = -\mu^a;\quad a'^a{}(0) = \pi n^a/s \ ,
\end{align}
where primes stand for derivatives with respect to the $r$ coordinate.

The inverse metric in the neighborhood of the horizon at zeroth order immediately follows:
\begin{align}\label{eq:sum_up_metric_0}
   g^{(0)rr} = -k(r);\quad g^{(0)r\mu} = u^\mu;\quad g^{(0)\mu\nu} = \frac{1}{f(r)} P^{\mu\nu} \ .
\end{align}

The corrections to the inverse metric are found as:
\begin{align}
 g^{(1)AB} = -g^{(0)AC}g^{(1)}_{CD}g^{(0)DB} \ .
\end{align}
This gives:
\begin{align}
\begin{split}\label{eq:sum_up_metric_1}
   g^{(1)rr} & = -u^\sigma g^{(1)}_{\sigma\delta} u^\delta \OnH 0\\
   \quad g^{(1)r\mu} & = -\frac{1}{f(r)} u^\sigma g^{(1)}_{\sigma\delta} P^{\delta\mu} \OnH 0\\
   \quad g^{(1)\mu\nu} & = -\frac{1}{f^2(r)} P^{\mu\sigma} g^{(1)}_{\sigma\delta} P^{\delta\nu}
   \OnH - \frac{1}{f^2(0)} h^{\mu\sigma} g^{(1)}_{\sigma\delta} h^{\delta\nu} \ ,
\end{split}
\end{align}
where ``$r=0$'' indicates the horizon value, and $h^{\mu\nu}$ is the 4d fluid (flat) metric.

The conventions for the Christoffel symbols and the Riemann are as follows:
\begin{align}
 \Gamma_{ABC} &= \del_{(B} g_{C)A} - \frac{1}{2}\del_A g_{BC} \\
 \Gamma^A_{BC} &= g^{AD}\Gamma_{DBC} \\
 R_{ABC}{}^D &= \del_B\Gamma^D_{AC} + \Gamma^D_{BE}\Gamma^E_{AC} - (A \leftrightarrow B) \ .
\end{align}

The non-vanishing lower Christoffel symbols at zeroth order read:
\begin{align}\label{eq:sum_lower_Christoffel_0}
\begin{split}
 \Gamma^{(0)}_{\mu r\nu} = \frac{1}{2}(f' P_{\mu\nu} + k' u_\mu u_\nu) ;\quad
 \Gamma^{(0)}_{r \mu \nu} = -\frac{1}{2}(f' P_{\mu\nu} + k' u_\mu u_\nu) \ .
 \end{split}
\end{align}

The non-vanishing upper Christoffel symbols at zeroth order are:
\begin{align}\label{eq:sum_upper_Christoffel_0}
\begin{split}
 \Gamma^{r(0)}_{r\mu} = -\frac{1}{2}k' u_\mu;\quad & \Gamma^{\mu(0)}_{r\nu} = \frac{f'}{2f}P^\mu_\nu;\quad
 \Gamma^{\mu(0)}_{\nu\rho} = -\frac{1}{2}u^\mu(f' P_{\nu\rho} + k' u_\nu u_\rho) \\
 & \Gamma^{r(0)}_{\mu\nu} = \frac{1}{2} k(r) (f' P_{\mu\nu} + k' u_\mu u_\nu) \OnH 0 \ .
 \end{split}
\end{align}

The Riemann tensor components of type $R_{\mu\nu A}{}^{B}$ read:
\begin{align}\label{eq:sum_R_munuA_B_0}
\begin{split}
 R^{(0)}_{\mu\nu r}{}^{r} = R^{(0)}_{\mu\nu \rho}{}^{r}& = 0;\quad
 R^{(0)}_{\mu\nu r}{}^{\rho} \OnH -\frac{f'k'}{4f}u_\mu P^\rho_\nu -(\mu \leftrightarrow \nu)\\
 &R^{(0)}_{\mu\nu\rho}{}^{\sigma} \OnH \frac{f'k'}{4}u_\mu u^\sigma P_{\nu\rho} -(\mu \leftrightarrow \nu) \ .
 \end{split}
\end{align}

The nonzero components of $R^{(0)}_{\mu\nu AB}$ are:
\begin{align}\label{eq:sum_R_0_lower}
 R^{(0)}_{\mu\nu r\rho} = -\frac{1}{4}f'k'\ell_\mu P_{\nu\rho} - (\mu\leftrightarrow\nu) \ .
\end{align}

The Riemann tensor components of type $R_{\mu\nu}{}^{AB}$ are:
\begin{align}\label{eq:sum_R_munu_AB_0}
\begin{split}
 R^{(0)}_{\mu\nu }{}^{\rho r} = 0;\quad
 R^{(0)}_{\mu\nu }{}^{\rho \sigma} \OnH \frac{f'k'}{4f} u_\mu (u^\sigma P^\rho_\nu - u^\rho P^\sigma_\nu) - (\mu \leftrightarrow \nu) \ .
 \end{split}
\end{align}

Some more Riemann components at zeroth order which we will need:
\begin{align}\label{eq:sum_R_others_0}
\begin{split}
 R^{(0)}_{\mu r \alpha}{}^{\epsilon} \OnH {} & -\frac{1}{2}u^\epsilon
 \left(P_{\mu\alpha} \left(f''-\frac{f'{}^2}{2f}\right) +u_\mu u_\alpha k''\right) +\frac{f'k'}{4f}P^\epsilon_\mu u_\alpha
 \\
 R^{(0)}_{\mu r}{}^{r \epsilon} \OnH {}&  g^{(0)r \beta} R^{(0)}_{\mu r \beta}{}^{\epsilon} =
 \frac{1}{2}u^\epsilon u_\mu k'' - \frac{f'k'}{4f}P^\epsilon_\mu \\
 R^{(0)}_{\mu r \alpha}{}^{r} \OnH {}&  g^{(0)r \beta} g^{(0)}_{r \rho} R^{(0)}_{\alpha \beta \mu}{}^{\rho} =
 \frac{f'k'}{4}P_{\alpha\mu} \ .
 \end{split}
\end{align}

The first-order lower Christoffel symbols are constructed from the derivatives $\del_\mu g^{(0)}_{AB}$ and $\del_r g^{(1)}_{AB}$. The nonzero components are:
\begin{align}\label{eq:sum_lower_Christoffel_1}
\begin{split}
 \Gamma^{(1)}_{\mu r\nu}  = & \del_{[\mu}u_{\nu]}+\frac{1}{2}g'^{(1)}_{\mu\nu}; \quad
 \Gamma^{(1)}_{r \mu \nu} = -\del_{(\mu}u_{\nu)}-\frac{1}{2}g'^{(1)}_{\mu\nu}\\
 \Gamma^{(1)}_{\mu \nu \rho} = & -\frac{1}{2}P_{\nu\rho}\del_\mu f +P_{\mu(\nu} \del_{\rho)} f
 -\frac{1}{2}u_\nu u_\rho \del_\mu k + u_\mu u_{(\nu} \del_{\rho)} k
 \\&\quad +(f+k)\left(-\del_\mu u_{(\nu} u_{\rho)} +\del_\rho u_{(\mu} u_{\nu)}
 +\del_\nu u_{(\mu} u_{\rho)} \right) \ .
 \end{split}
\end{align}

The first-order upper Christoffel symbols are found as:
\begin{align}
 \Gamma^{A(1)}_{BC} = g^{(1)AD}\Gamma^{(0)}_{DBC} + g^{(0)AD}\Gamma^{(1)}_{DBC} \ .
\end{align}
The nonzero components read:
\begin{align}\label{eq:sum_upper_Christoffel_1}
\begin{split}
 \Gamma^{r(1)}_{r\mu} = {} &  \frac{1}{2}a_\mu +\frac{1}{2}u^\nu g'^{(1)}_{\mu\nu}
 -\frac{f'}{2f}u^\rho g^{(1)}_{\rho\delta}P^\delta_\mu\OnH
 \frac{1}{2} \left(a_\mu +u^\nu g'^{(1)}_{\mu\nu} \right)
 \end{split}\\
 \begin{split}
 \Gamma^{\mu(1)}_{r\nu} = {} & \frac{1}{f} \left(P^{\mu\rho} \del_{[\rho} u_{\nu]}
 +\frac{1}{2} P^{\mu\rho} g'^{(1)}_{\rho\nu}
 -\frac{f'}{2f} P^{\mu\rho} g^{(1)}_{\rho\delta} P^\delta_\nu \right)
 \end{split} \\
 \begin{split}
 \Gamma^{r(1)}_{\mu\nu} = {} & \frac{1}{2} k g'^{(1)}_{\mu\nu}
 +\frac{1}{2} u^\sigma g^{(1)}_{\sigma\delta} u^\delta \left(f'P_{\mu\nu}+k'u_\mu u_\nu\right)
 -\frac{1}{2}\left(P_{\mu\nu} u^\beta \del_\beta f  + u_\mu u_\nu u^\beta\del_\beta k   \right)\\
 & -u_{(\mu}\del_{\nu)} k - f P_\mu^\rho P_\nu^\sigma \del_{(\rho} u_{\sigma)} -
 k a_{(\mu} u_{\nu)}  \OnH
 -\frac{1}{2}P_{\mu\nu} u^\rho \del_\rho f  -
 fP_\mu^\rho P_\nu^\sigma \del_{(\rho} u_{\sigma)}
 \end{split}\\
 \begin{split}
 \Gamma^{\mu(1)}_{\nu\rho} = {} &
 \frac{1}{f}\left( -\frac{1}{2}P_{\nu\rho} P^{\mu\beta}\del_\beta f + P^\mu_{(\nu}\del_{\rho)} f -\frac{1}{2}u_\nu u_\rho P^{\mu\alpha} \del_\alpha k\right) \\
 & +\left(1+\frac{k}{f} \right) \left(-P^{\mu\alpha}\del_\alpha u_{(\nu}u_{\rho)}
 +u_{(\nu}\del_{\rho)} u^\mu\right)
 -u^\mu\left(\del_{(\nu} u_{\rho)} +\frac{1}{2} g'^{(1)}_{\nu\rho} \right) \\
 & +\frac{1}{2f}u^\rho g^{(1)}_{\rho\delta} P^{\delta\mu} \left(f' P_{\nu\rho} +k'u_\nu u_\rho \right) \OnH
 \frac{1}{f}\left(P^\mu_{(\nu} \del_{\rho)} f -\frac{1}{2}P_{\nu\rho} P^{\mu\sigma} \del_\sigma f\right) \\ &
 +u_{(\nu} \del_{\rho)} u^\mu -  \del^\mu u_{(\nu}u_{\rho)}
 -u^\mu \left( P^\sigma_\nu P^\lambda_\rho \del_{(\sigma} u_{\lambda)}
 +\frac{1}{2} g'^{(1)}_{\nu\rho}\right) \ ,
 \end{split}
\end{align}
where
$$a^\mu \equiv u^\nu\del_\nu u^\mu$$
is the acceleration vector.

First-order Riemann components whose $r$ dependence is needed:
\begin{align}\label{eq:sum_Riemann_r_dep_1}
\begin{split}
 R^{(1)}_{\mu\nu r}{}^r = {} &  \frac{1}{2} \del_\mu(k'u_\nu) +\frac{f'}{2f}
 \left( k \del_{[\nu} u_{\mu]} -\frac{1}{2}u_\nu \del_\mu k \right)
 -(\mu \leftrightarrow \nu) \OnH \frac{1}{2} \del_\mu(k'u_\nu) -(\mu \leftrightarrow \nu)
 \end{split} \\
 \begin{split}
 R^{(1)}_{\mu\nu \rho}{}^r = {} & \frac{1}{4} \del_\nu k \left( f'P_{\mu\rho} +k' u_\mu u_\rho \right)
 +\frac{1}{2} k \left(\del_\nu f' P_{\mu\rho} +\del_\nu k' u_\mu u_\rho\right)
 +k(f'+k')\del_\nu u_{(\mu} u_{\rho)} \\&
 -\frac{1}{2} k' u_\nu \left(-\frac{1}{2}u^\sigma\del_\sigma f P_{\mu\rho}
 -\frac{1}{2} u_\rho \del_\mu k - f P^\sigma_\rho \del_{(\mu} u_{\sigma)}
 -\frac{1}{2}k a_\mu u_\rho \right)\\&
 +\frac{kf'}{2f}\left(-\frac{1}{2}P_{\mu\rho}P_\nu^\delta \del_\delta f +\frac{1}{2}P_{\nu\rho}\del_\mu f
 -\frac{1}{2}u_\mu u_\rho \del_\nu k \right)\\&
 +\frac{kf'}{2} \left(1+\frac{k}{f}\right) \left(-P_\nu^\delta\del_\delta u_{(\mu}u_{\rho)}
 +u_{(\mu}\del_{\rho)} u_\nu \right) +\frac{kk'}{2}u_\nu\del_{(\mu}u_{\rho)} - (\mu \leftrightarrow \nu) \\
 \OnH {}& -\frac{1}{2}k' u_\mu\left(\frac{1}{2}P_{\nu\rho}u^\sigma\del_\sigma f + f P_\nu^\lambda P_\rho^\sigma\del_{(\lambda}u_{\sigma)} \right)
      - (\mu\leftrightarrow\nu)
 \end{split} \\
 \begin{split}
 R^{(1)}_{\mu\nu}{}^{\rho r} = {} & \frac{1}{2} u^\rho u_\nu \del_\mu k' +
 \left(k'-\frac{kf'}{f}\right)
        \left[\frac{1}{2}u^\rho \del_\mu u_\nu +\frac{1}{4} \left(1-\frac{k}{f}\right) u_\nu \del_\mu u^\rho \right.
        \\ &
        \left. +\frac{1}{4} \left(1+\frac{k}{f}\right) u_\nu P^{\rho \sigma} \del_\sigma u_\mu +\frac{1}{4f} P^\rho_\mu P_\nu^\delta \del_\delta f \right] -\frac{1}{4f} P^\rho_\mu \left( k'+\frac{kf'}{f} \right) \del_\nu f
        \\ {} &
        - \frac{f'}{4f} \del_\mu k (u^\rho u_\nu +P_\nu^\rho) + \frac{k}{2f}\del_\nu f' \ P_\mu^\rho
        - (\mu\leftrightarrow\nu) \\
        \OnH {} & \frac{1}{2}u^\rho \del_\mu (k'u_\nu) +
        \frac{k'}{4} u_\nu
                \left[2P^{\rho\sigma} \del_{(\sigma} u_{\mu)} +\frac{1}{f} P_\mu^\rho u^\delta \del_\delta f \right]
        - (\mu\leftrightarrow\nu) \ .
 \end{split}
\end{align}

For the other needed Riemann components (whose $r$ dependence is not needed), we have:
\begin{align}
 \begin{split}\label{eq:sum_Riemann_1_no_r_dep_1}
   R^{(1)}_{\mu\nu r}{}^\rho \OnH{}&
   \frac{1}{4f}u_\mu \biggr\{f'\left(\frac{1}{f}P_\nu^\rho u^\sigma\del_\sigma f + 2P^{\rho\sigma}\del_{(\sigma} u_{\nu)}\right)
      \\ & \qquad\quad \quad
      - k'\left(2P^{\rho\sigma}\del_{[\sigma}u_{\nu]} + P^{\rho\sigma}g'^{(1)}_{\sigma\nu}
      - \frac{f'}{f} P^{\rho\sigma} g^{(1)}_{\sigma\nu}\right)\biggr\} \\
      & + \frac{1}{2}P_\mu^\rho\left(\del_\nu\frac{f'}{f} - \frac{f'}{2f}(a_\nu + g'^{(1)}_{\nu\sigma}u^\sigma) \right)
      - (\mu\leftrightarrow\nu)
  \end{split}\\
  \begin{split}\label{eq:sum_Riemann_1_no_r_dep_2}
   R^{(1)}_{\mu\nu\rho}{}^\sigma \OnH {}& -\frac{1}{4}(f'P_{\mu\rho} + k'u_\mu u_\rho)\left(\frac{1}{f}P_\nu^\sigma u^\lambda\del_\lambda f
      + P_\nu^\lambda\del_\lambda u^\sigma + \del^\sigma u_\nu - u^\sigma g'^{(1)}_{\nu\lambda}u^\lambda \right) \\
      &- \frac{u^\sigma u_\mu}{2} \biggr\{\sqrt{f}P_{\nu\rho} u^\lambda\del_\lambda\frac{f'}{\sqrt{f}}+u_\rho\del_\nu k'
          + k'\left(\del_{[\nu}u_{\rho]} - \frac{g'^{(1)}_{\nu\rho}}{2}\right) +f' P_\rho^\lambda \del_{(\lambda}u_{\nu)} \biggr\}
           \\  &
          + \frac{1}{2}u^\sigma\left(k'u_\rho\del_\mu u_\nu - f P_{\mu\rho}P_\nu^\lambda\del_\lambda\frac{f'}{f} \right)\\&
      + \frac{1}{2}f' P_\mu^\sigma\left(\frac{1}{2f}P_{\nu\rho}u^\lambda\del_\lambda f + P_\nu^\kappa P_\rho^\lambda\del_{(\kappa}u_{\lambda)} \right)
      - (\mu\leftrightarrow\nu)
  \end{split}\\
 \begin{split}\label{eq:sum_Riemann_1_no_r_dep_3}
   R^{(1)}_{\mu r \alpha}{}^r \OnH{}& g^{r\beta(0)} g^{(0)}_{r\rho} R^{(1)}_{\alpha\beta\mu}{}^{\rho} =
   -\frac{1}{2} u_\mu P_{\alpha\lambda}\del^\lambda k' +
   \frac{k'}{4} \left(\del_\mu u_\alpha- u_\mu a_\alpha  - P_{\alpha\lambda}\del^\lambda u_\mu\right)
   \\ & -\frac{f'}{2}P_{\lambda(\mu}\del^\lambda u_{\alpha)}
   -\frac{1}{2}\sqrt{f} P_{\alpha\mu} u^\lambda\del_\lambda \left(\frac{f'}{\sqrt{f}}\right)
   +\frac{k'}{4} \left(u_\mu u^\lambda g'^{(1)}_{\lambda\alpha} + P_\alpha^\lambda g'^{(1)}_{\lambda\mu} \right)
   \\&
   +\frac{1}{4}\left[f'P_{\mu\alpha} + k' u_\mu u_\alpha \right]u^\delta g'^{(1)}_{\delta\lambda}u^\lambda \ .
 \end{split}
\end{align}
Sometimes it's useful to decompose the last two indices of $R^{(1)}_{\mu\nu\rho}{}^\sigma$ along and orthogonally to $u^\mu$:
\begin{align}
 \begin{split}\label{eq:sum_Riemann_1_decomp}
   R^{(1)}_{\mu\nu\rho}{}^\sigma u^\rho u_\sigma \OnH {}& \frac{1}{2}\del_\mu(k' u_\nu) - (\mu\leftrightarrow\nu) \\
   R^{(1)}_{\mu\nu\rho}{}^\beta u^\rho P_\beta^\sigma \OnH {}& \frac{1}{2}k'u_\mu\left(\frac{1}{2f}P_\nu^\sigma u^\lambda\del_\lambda f
      + P^{\sigma\lambda}\del_{(\lambda}u_{\nu)} \right) - (\mu\leftrightarrow\nu) \\
   R^{(1)}_{\mu\nu\alpha}{}^\sigma P_\rho^\alpha u_\sigma \OnH {}& \frac{1}{2}u_\mu\left(\sqrt{f}P_{\nu\rho}u^\lambda\del_\lambda\frac{f'}{\sqrt{f}} +
      f' P^\lambda_\rho\del_{(\lambda}u_{\nu)} + k'\left(P^\lambda_\rho\del_{[\nu}u_{\lambda]} - \frac{1}{2}g'^{(1)}_{\nu\lambda}P^\lambda_\rho \right)\right) \\
      &+ \frac{1}{2}P_{\mu\rho}\left(f P_\nu^\lambda\del_\lambda\frac{f'}{f} - \frac{1}{2}f'(a_\nu + g'^{(1)}_{\nu\lambda} u^\lambda)\right) - (\mu\leftrightarrow\nu) \\
   R^{(1)}_{\mu\nu\alpha}{}^\beta P^\alpha_\rho P_\beta^\sigma \OnH {}& \frac{1}{2}f'\left(
        P_\mu^\sigma\left(\frac{1}{f} P_{\nu\rho}u^\lambda\del_\lambda f + P_\nu^\kappa P_\rho^\lambda\del_{(\kappa}u_{\lambda)} \right)
        - P_{\mu\rho}P_\nu^\kappa P^{\sigma\lambda}\del_{(\kappa}u_{\lambda)} \right)\\&
         - (\mu\leftrightarrow\nu) \ .
 \end{split}
\end{align}

As for the field strength tensor, all we will need is:
\begin{align}
 \begin{split}
   & A_r^{a(0)} = A_r^{a(1)}=0; \quad A_\mu^{a(0)} = a^a(r,x^\mu)u_\mu\\
   & F_{\kappa r}^{a(0)}=-a'^a u_\kappa; \quad F_{\mu\nu}^{a(0)} = 0 \\
   & F_{\kappa r}^{a(1)}=-A^{a}_\kappa{}'^{(1)}; \quad F_{\mu\nu}^{a(1)} = \del_\mu(a^au_\nu)-\del_\nu(a^au_\mu) \\
   & F_{\mu\nu}^{a(2)} = \del_\mu(A^a_\nu{}^{(1)})-\del_\nu(A^a_\mu{}^{(1)}) \ .
 \end{split}
\end{align}

\end{document}